\documentclass[12pt, preprint]{aastex}

\usepackage{natbib}

\newcommand{\ug}{u-g}
\newcommand{\gr}{g-r}
\newcommand{\jbol}{j_{\rm bol}}
\newcommand{\Zsun}{Z_{\odot}}

\newcommand{\etastar}{\eta_\star}
\newcommand{\EBL}{I_{\rm EBL}}
\newcommand{\eblunit}{{\rm nW\,m^{-2}\,sr^{-1}}} 

\newcommand{\rhoz}{\rho_Z}
\newcommand{\rhos}{\rho_{\star}}
\newcommand{\rhosb}{\rho_{\star,b}}
\newcommand{\rhosd}{\rho_{\star,d}}
\newcommand{\rhocrit}{\rho_{\rm crit}}

\newcommand{\rhouv}{\rho_{\rm UV}}
\newcommand{\Oz}{\Omega_Z}
\newcommand{\Os}{\Omega_{\star}}
\newcommand{\Om}{\Omega_{\rm M}}
\newcommand{\Ol}{\Omega_{\Lambda}}
\newcommand{\Ob}{\Omega_b}

\def\avg#1{\left\langle#1\right\rangle}
\def\rp#1{\left(#1\right)}

\newcommand{\lam}{\lambda}
\newcommand{\Lam}{\Lambda}

\newcommand{\hinv}{{h^{-1}}}
\newcommand{\mpc}{{\rm\,Mpc}}

\newcommand{\himpc}{\hinv{\rm\,Mpc}}
\newcommand{\hikpc}{\hinv{\rm\,kpc}}
\newcommand{\kms}{{\rm\,km\ s^{-1}}}
\newcommand{\kmsmpc}{{\rm\ km\ s^{-1}\ Mpc^{-1}}}
\newcommand{\Gyr}{{\rm\,Gyr}}
\newcommand{\yr}{{\rm yr}}
\newcommand{\Msun}{M_{\odot}}
\newcommand{\himsun}{\hinv{\Msun}}
\newcommand{\ltsim}{\lesssim}

\newcommand{\Mstar}{M_\star}

\newcommand{\dd}{{\rm d}}
\newcommand{\erg}{{\rm erg}}
\newcommand{\s}{{\rm s}}

\newcommand{\Hz}{{\rm Hz}}

\slugcomment{Submitted to ApJ}

\shorttitle{The History of Cosmological Star Formation}
\shortauthors{Nagamine, Ostriker, Cen, \& Fukugita}

\begin{document}

\title{The History of Cosmological Star Formation: Three Independent Approaches and a Critical Test Using the Extragalactic Background Light}

\author{Kentaro Nagamine\altaffilmark{1}, Jeremiah P. Ostriker\altaffilmark{2}, Masataka Fukugita\altaffilmark{3,4}, \& Renyue Cen\altaffilmark{2}}

\altaffiltext{1}{University of California, San Diego, Center for Astrophysics and Space Sciences, CA 92093-0424, U.S.A., Email: knagamine@ucsd.edu} 

\altaffiltext{2}{Princeton University Observatory, Princeton, NJ 08544, U.S.A.}

\altaffiltext{3}{Institute for Cosmic Ray Research, University of Tokyo, 
Kashiwa 2778582, Japan}

\altaffiltext{4}{Institute for Advanced Study, Einstein Drive, Princeton NJ, 08540, U.S.A.}


\begin{abstract}

Taking three independent approaches, we investigate the 
simultaneous constraints set on the cosmic star formation history 
from various observations, including stellar mass density and 
extragalactic background light (EBL).  We compare results based on: 
1) direct observations of past light-cone, 
2) a model using local fossil evidence constrained by 
SDSS observations at $z\sim 0$ (the `Fossil' model), and 
3) theoretical {\it ab initio} models from 
three calculations of cosmic star formation history: 
(a) new (1024)$^3$ Total Variation Diminishing (TVD) 
cosmological hydrodynamic simulation, 
(b) analytic expression of Hernquist \& Springel based on 
cosmological Smoothed Particle Hydrodynamics (SPH) simulations, 
and (c) semi-analytic model of Cole et al.
We find good agreement among the three independent approaches
up to the order of observational errors, except that 
all the models predict bolometric EBL of 
$I_{\rm tot}\simeq 37-52$ nW m$^{-2}$ sr$^{-1}$, 
which is at the lower edge of the the observational 
estimate by \citet{Hauser01}. 
We emphasize that the Fossil model that consists of 
two components --- spheroids and disks ---, when normalized
to the local observations, provides a surprisingly simple 
but accurate description of the cosmic star formation history 
and other observable quantities.
Our analysis suggests that the consensus global parameters 
at $z=0$ are: 
$\Os = 0.0023\pm 0.0004$, $\EBL = 43\pm 7\,\eblunit$, 
$\dot{\rhos}=(1.06\pm 0.22)\times 10^{-2}\,{\rm \Msun}\,\yr^{-1}\,\mpc^{-3}$, 
$\jbol = (3.1\pm 0.2)\times 10^8\,L_\odot\,\mpc^{-3}$. 
\end{abstract}



\section{Introduction}
\label{section:intro}

Targets of cosmological investigations during the last decade have 
largely shifted from the global, geometrical properties of 
the universe to the detailed contents thereof, 
particularly the origin and evolution of structure from 
smooth initial conditions.  Of the components added since 
the decoupling of matter and radiation, stars are the 
most easily observable feature (emitting the most electromagnetic 
energy) and of the greatest historical significance. 
Approximately $6\pm2$\% \citep[e.g.,][]{Fukugita04} of the baryons 
of the universe have condensed out into stars at the current epoch, 
and the time history of the process whereby this occurred is 
the subject of intense study at present. 
What is not widely realized is that there are three independent 
approaches by which this study can be pursued:

\begin{enumerate}
\item We can use the universe directly as a time-machine:
in areas of the sky like the Hubble Deep Field \citep{Williams96}, 
attempts have been made to analyze the past light-cone.
The rate of transformation of gas into stars is measured as 
a function of the look-back time or redshift,
which produces what has been called the ``Madau Diagram'' 
\citep{Madau96}.  There are three major uncertainties 
involved in this diagram: 
(1) dust obscuration; 
(2) relation between emission and star formation rate (SFR),
which includes the uncertainty in the initial mass function (IMF); 
(3) faint-end slope of the luminosity function, especially 
at high-redshift. All of these three contribute roughly on the 
same order. 
Recent observational estimates at $z\gtrsim 4$ suggest a 
decline in SFR with increasing redshift \citep{Iwata03, Bunker04, 
Bouwens04a, Giavalisco04, Ouchi04a, Bouwens05}, but due to 
uncertainties in dust extinction for ultra-violet (UV) light
and the faint-end slope of the luminosity function, 
the trend does not seem to be inconsistent with 
being constant at $z>3$ \citep[See also Figure~2 of][]{Nachos1}.

\item The second method predated the direct method outlined above. 
It uses the fossil record from our own and other galaxies 
to catalog stars of different ages and to, by an essentially 
demographic investigation, unravel the history of star formation 
in the local Universe using the theory of stellar evolution.
When direct age estimates are not available, as is the case 
for external galaxies, color distributions can provide 
a useful proxy to their history, distinguishing between
systems with and without ongoing star formation activity, 
although there are degeneracies among age, 
amount of stars formed, metallicity, and IMF. 
This method was pioneered by \citet*{Searle73}. 
Here one makes a simple ansatz for the star formation history 
--- chosen in the quoted paper to be as a simple 
declining exponential --- 
and an IMF, and then computes, via the standard theories of 
stellar evolution and stellar atmospheres, the evolving spectral 
output from the assemblage of stars. 
They found that one could easily and plausibly simulate the 
observed colors of different types of stellar systems. 
We will update the quantitative input to this theory, 
leaving the essential basis unchanged, and find that 
this ``Fossil'' model fits with uncanny accuracy (given the 
simplicity of the method) modern observations as obtained 
by the Sloan Digital Sky Survey \citep[SDSS;][]{York00}. 
We will then use SDSS results to determine the parameters 
of the Fossil model.

\item The third method is to attempt to solve the problem 
by {\it ab initio} theory. One takes the initial conditions 
from the observed cosmic microwave background radiation, 
adopts a favorite cosmological model, 
e.g., the standard concordance $\Lam$ cold dark matter (CDM) model 
\citep{Ostriker95, Perlmutter98, Riess98, Spergel03, Tegmark04}, 
combines this with the standard physical equations of gravity, 
hydrodynamics, atomic physics, radiative transfer, {\it etc.}, 
and simulates the history of the universe by numerical integration, 
obtaining as the most important by-product 
the history of star formation 
\citep[e.g.,][]{Nag00, Nag01a, Ascasibar02, SH03b, Nachos1, Nag06a}.
One could also employ the so-called semi-analytic models of 
galaxy formation \citep[e.g.,][]{Kau93, Cole00, Som01}, 
where a dark-matter halo merger-tree, either constructed 
using the extended Press-Schechter formalism 
\citep{Bond91, Lacey93} or N-body simulations, is supplemented 
with a set of equations to model transformation of gas into stars 
within virialized dark matter halos. 

\end{enumerate}

If the observational errors are too significant
or the theoretical modeling in error, 
these three independent approaches might not agree with 
one another when they should.  
We need to pay particular attention to the errors 
that are involved in various observations. 
The quantities at near zero redshift are usually measured 
without too serious systematic errors. 
On the other hand, high redshift observations, 
such as those of star formation rates are subject to 
large systematic uncertainties from the obscuration and 
the faint-end slope of luminosity functions. 
Therefore, we take the attitude that we constrain 
the three approaches with zero-redshift quantities, 
while high-redshift observations are used to examine 
the consistency within allowed errors. 
We also note that there are some uncertainty such as IMF, 
but these uncertainties are known and we can see 
how they propagate to our calculations or even constrain
those uncertainties. With understanding those uncertainties, 
we can conclude that the agreement of the three approaches 
is in fact good.
In addition, there is an independent check of all of these approaches.  
Each approach predicts two numbers: (1) the local bolometric 
electromagnetic energy output per unit volume 
$\jbol(z=0)$ [erg sec$^{-1}$ cm$^{-3}$] and (2) the 
local energy density $U(z=0)$ [erg cm$^{-3}$], which 
is measured as the extragalactic background light (EBL).
The first number $\jbol(z=0)$ provides a measure of the current 
star formation rate, while $U(z=0)$ constrains the integral of all
past star formation. The local energy output provides the normalization 
of the first two quoted approaches and consequently cannot be used as 
an independent check, but the EBL provides an 
independent test of the three approaches.

The paper is organized as follows. 
In Section~\ref{sec:fossil} we describe the fossil approach characterized 
by delayed exponential decay models of cosmological star formation and 
investigate the resulting color and mean age as a function of 
characteristic decay time and metallicity. 
As a result, we construct the `Fossil' model using two components, 
bulge and disk, based on the observed colors of galaxies. 
In Section~\ref{sec:sfr} we consider the star formation history,
as predicted by the Fossil model and other {\it ab initio}
theoretical models, and compare it with what is inferred from observation.
In Section~\ref{sec:omegastar} we discuss the stellar 
mass density, a time integrated version of the star formation history,
which is taken as a consistency test.
We then compute the EBL in Section~\ref{sec:ebl}
to study whether what is expected from the star formation history 
is consistent with observations. 
Conclusions are given in Section~\ref{sec:discussion}.
When necessary, we adopt a cosmological model with parameters 
$(\Om, \Ol, \Ob, h, n, \sigma_8)= (0.3, 0.7, 0.04, 0.7, 1.0, 0.9)$, 
where $h = H_0 / (100 \kmsmpc)$ is the Hubble parameter. 
Where $h$ is not explicit, $h=0.7$ is assumed.


\section{The Fossil model}
\label{sec:fossil}

\citet*{Searle73} showed that the colors of galaxies can be 
reproduced well by simple exponentially decaying star formation 
histories, $\dot{\rho}_* = A\, \exp(-t/\tau)$, where $\tau$ 
is the characteristic decay time. Here we revisit their model 
with a modern perspective. 
Throughout this paper, we take $t=0$ to be the epoch of the Big Bang.

The original model of \citet{Searle73} has the disadvantage 
that one has to assume the onset time of star formation by hand, 
increasing the number of free parameters. 
Therefore we adopt the delayed exponential model, 
\begin{equation}
\dot{\rho}_* = A\, (t/\tau) \exp(-t/\tau).
\label{eq:delayexp}
\end{equation}  
This model has a desirable feature that the SFR vanishes at $t=0$, 
and it has a peak at $t=\tau$ followed by an exponential tail 
with a characteristic time-scale $\tau$. 
The model with a very small value of $\tau$ resembles 
the instantaneous burst at $t=0$, and the one with $\tau \sim t_H$
(Hubble time) resembles the model with a constant SFR.
The delayed exponential model has the same minimal number of 
free parameters as the \citet{Searle73} model, 
a time-scale $\tau$ and a normalization $A$. 
Good agreement with other approaches, as we will show later, 
justifies the use of this functional form for the Fossil model. 

\begin{figure}
\begin{center}
\resizebox{9.2cm}{!}{\includegraphics{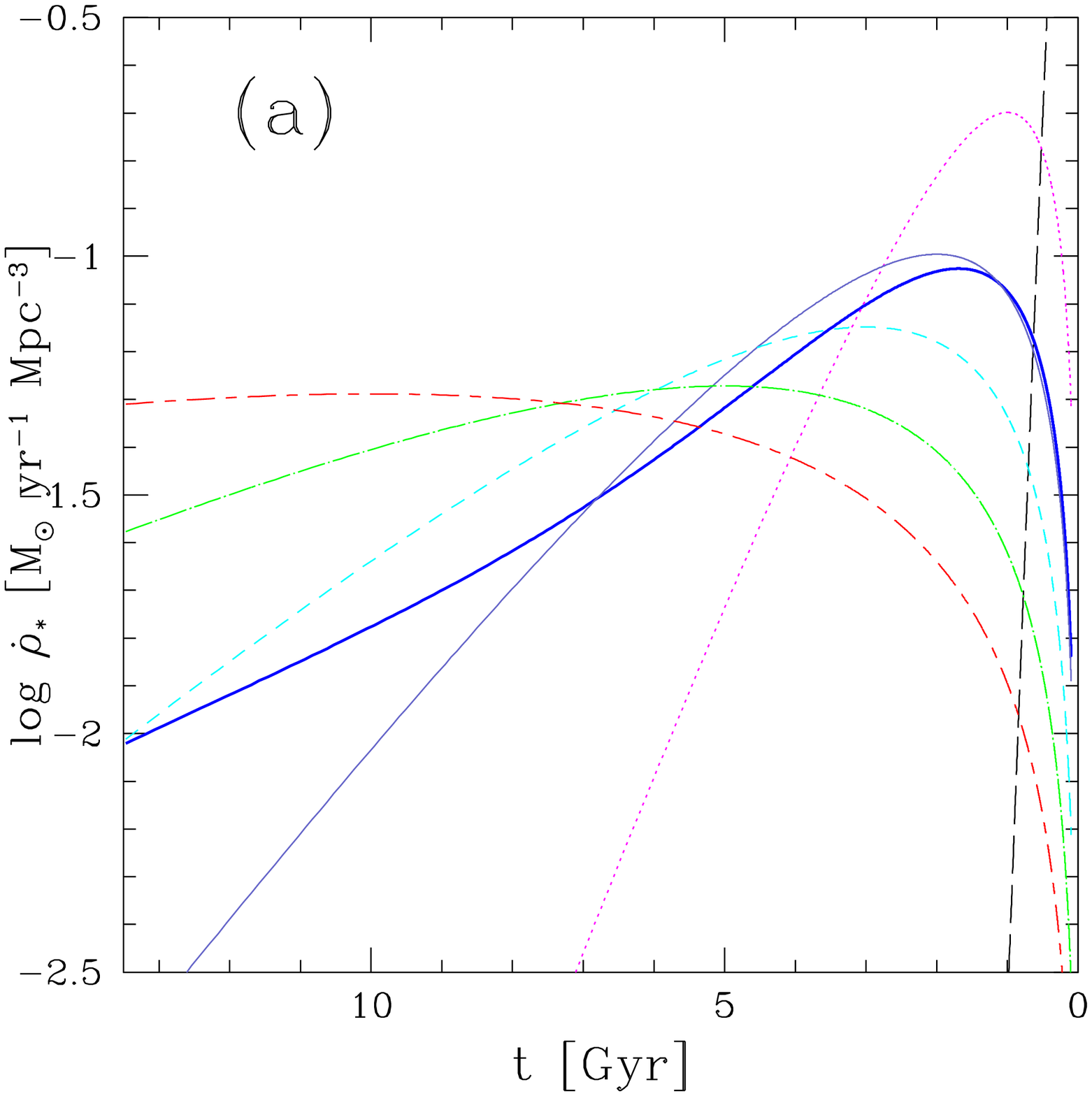}}\\
\vspace{0.2cm}
\resizebox{9.2cm}{!}{\includegraphics{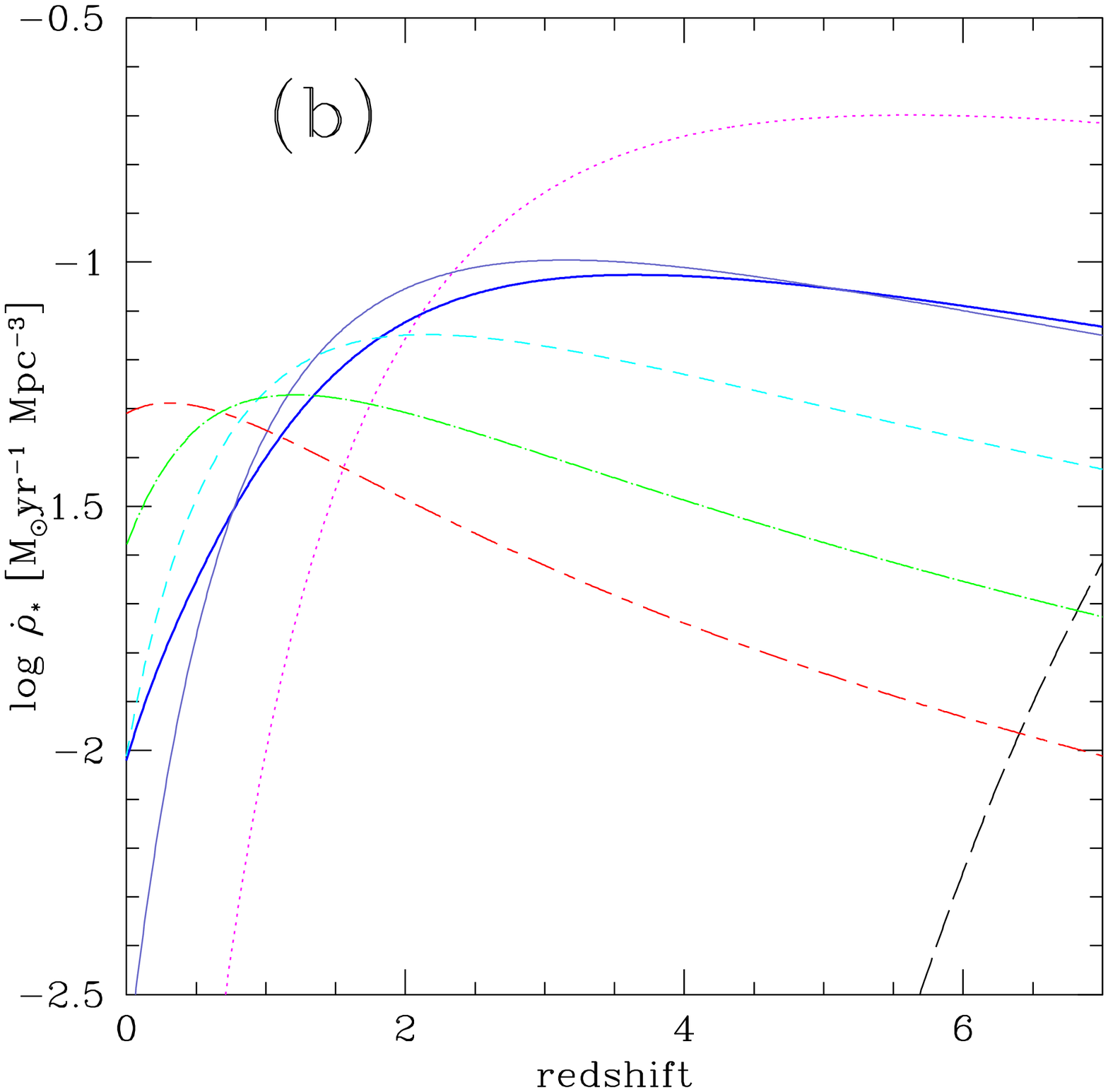}}%
\vspace{0.1cm}
\caption{Star formation histories of delayed exponential decay model
as functions of cosmic time ({\it panel a}) and redshift 
({\it panel b}). The normalization 
is fixed to equation~(\ref{eq:norm}) for each model.
The lines are for $\tau = 0.1$\,Gyr (black long-dashed), 1\,Gyr 
(magenta dotted), 2\,Gyr (blue solid), 3\,Gyr (cyan dashed), 
5\,Gyr (green dot-dashed), and 10\,Gyrs (red long-short-dashed). 
The blue thick solid line is the composite two population model 
with $\tau=1.5$\,Gyr and 4.5\,Gyr, which is taken as our
`Fossil' model. 
}
\label{fig:sfr_tau}
\end{center}
\end{figure}

Figure~\ref{fig:sfr_tau} shows the delayed exponential models 
of star formation history as functions of cosmic time 
(panel [$a$]) and redshift (panel [$b$]) for 
different values of $\tau$ as described in the caption.  
The normalization $A$ is adjusted so that the integral of the SFR, 
\begin{equation}
\int_0^{t_0} \dot{\rho_*} dt / \rhocrit = 0.004,
\label{eq:norm}
\end{equation}
where $\rhocrit$ is the critical mass density and 
$t_0$ is the current age of the universe. 
The {\it thick blue solid} line is a composite 
two population model with $\tau = 1.5$\,Gyr and 4.5\,Gyr 
which will be described below.

\begin{figure}
\epsscale{1.0}
\plotone{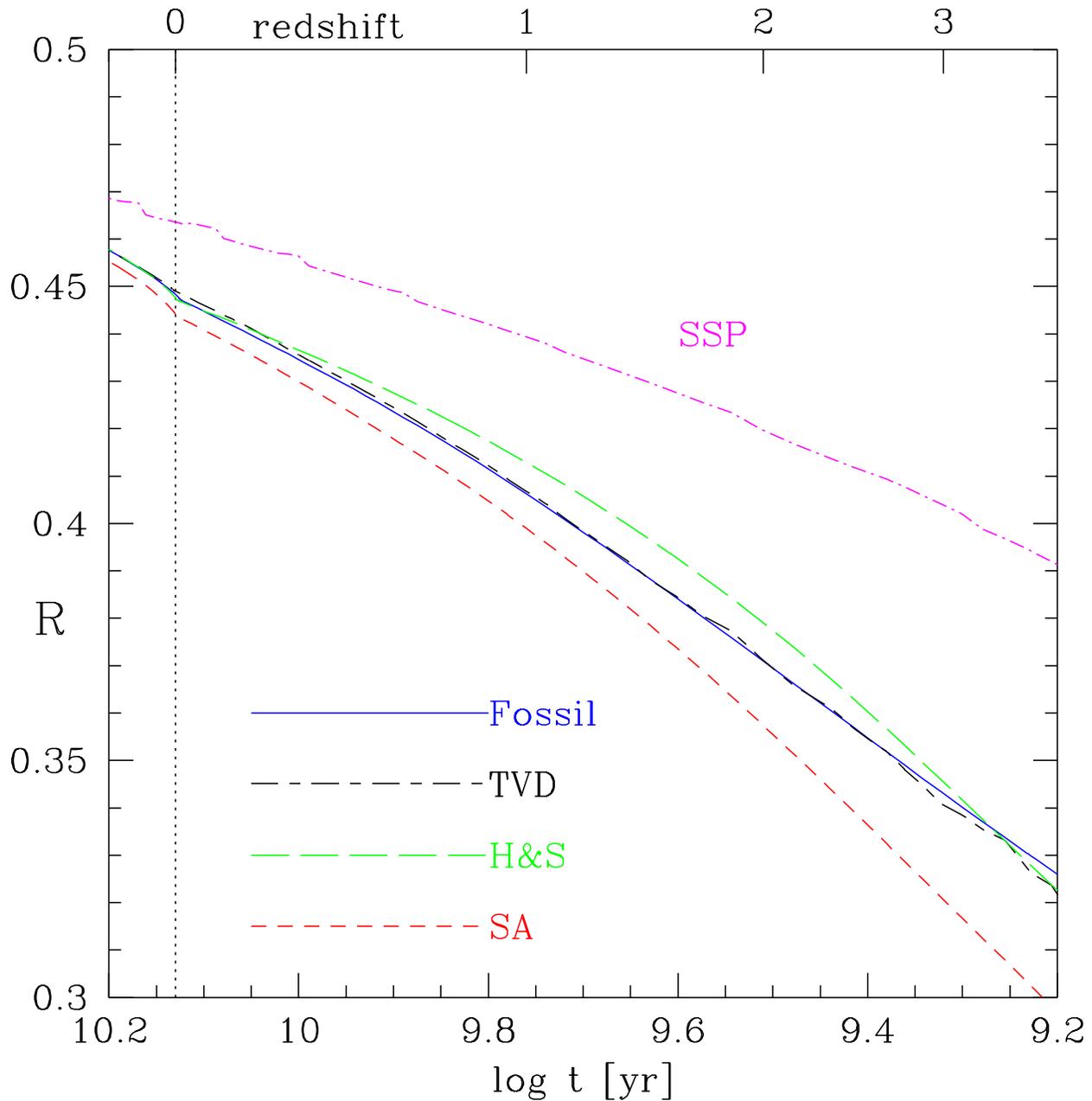}
\caption{Gas recycling fraction $R$ as a function of 
cosmic time (and redshift), computed using BClib03 
for solar metallicity and the Chabrier IMF ($0.01 - 100\,\Msun$). 
`SSP' stands for `Simple Stellar Population' with 
an instantaneous burst at $t=0$. 
}
\label{fig:recyc}
\end{figure}

We note that the integral of SFR (Equation~[\ref{eq:norm}]) 
is not equal to the stellar mass density observed today
because of gas recycling, i.e., gas lost by aging stars. 
The recycling fraction $R$, defined to be the ratio of 
the amount of recycled gas to the total amount of gas 
that was initially converted into stars, depends on 
the IMF and is a function of time.  
We show the recycling fraction in  Figure~\ref{fig:recyc}
computed using the population synthesis model of 
\citet[][hereafter BClib03]{BClib03}. 
This shows $R\simeq 0.45$ at $t=13.5$\,Gyr for the 
\citet{Chab03a} IMF integrated from 0.01 to 100\,$\Msun$\footnote{
BClib03 integrates the Chabrier IMF from
0.1 to 100$M_\odot$. We applied the correction for 
brown dwarfs that reduces $R$ from BClib03 
by a factor of 0.94.}.
In other words, approximately 55\% of the initial gas 
remains in stars and stellar remnants
(45\% in shining stars and $10$\% in remnants, i.e., 
white dwarfs, neutron stars and black holes, 
for which BClib03 assumes the mass, 
$0.55, 1.4$, and $2.0\,\Msun$, respectively.)
Therefore the normalization given by Equation~(\ref{eq:norm})
corresponds to the stellar mass density 
$\Os(z=0) \approx 0.0022$.  
We note that the recycling fraction of the SSP for the Salpeter IMF 
($0.1 - 100\,\Msun$) is $R=0.32$ at $t=13.5$\,Gyr.

\begin{figure}
\begin{center}
\resizebox{9cm}{!}{\includegraphics{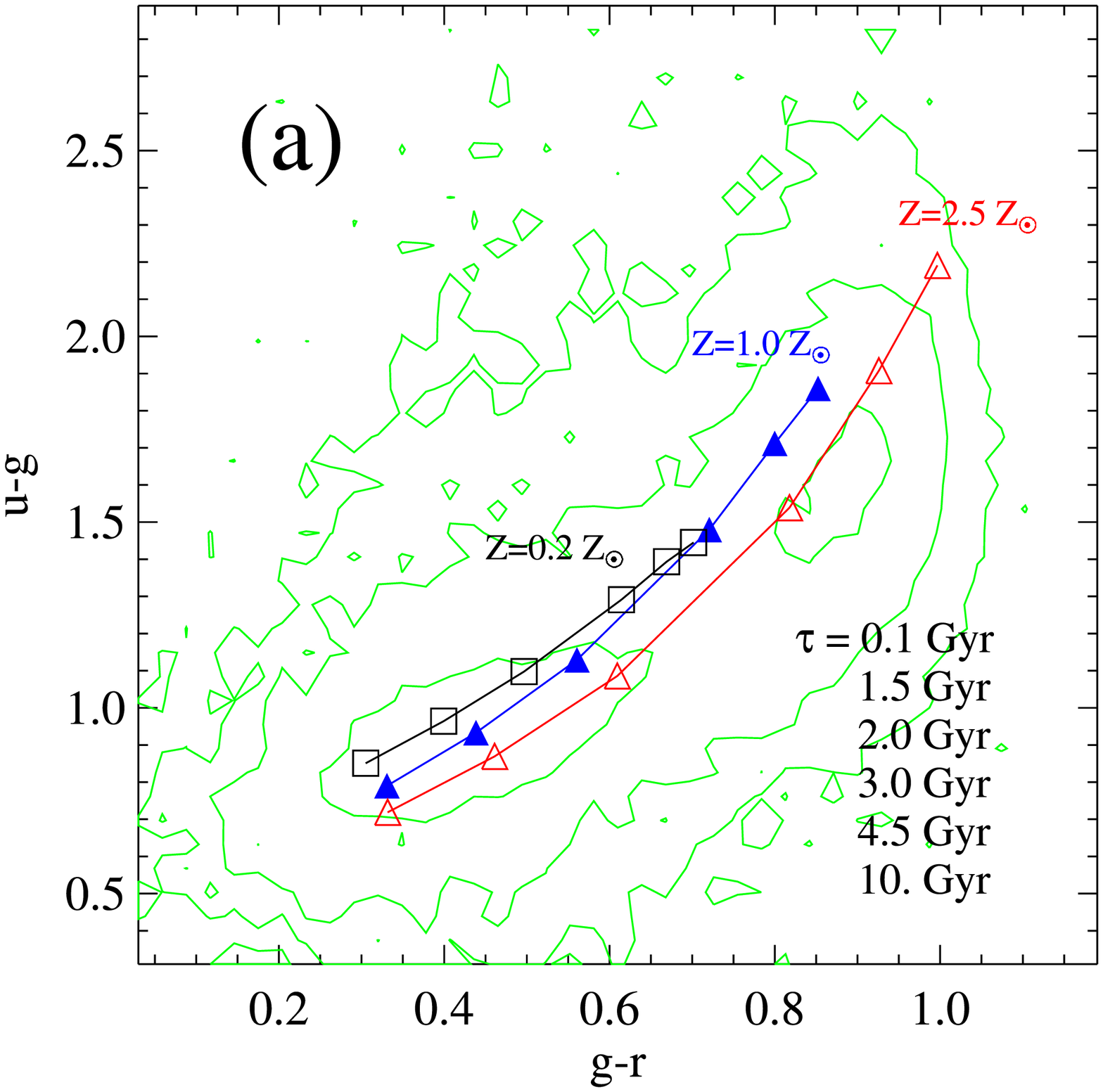}}\\
\resizebox{9cm}{!}{\includegraphics{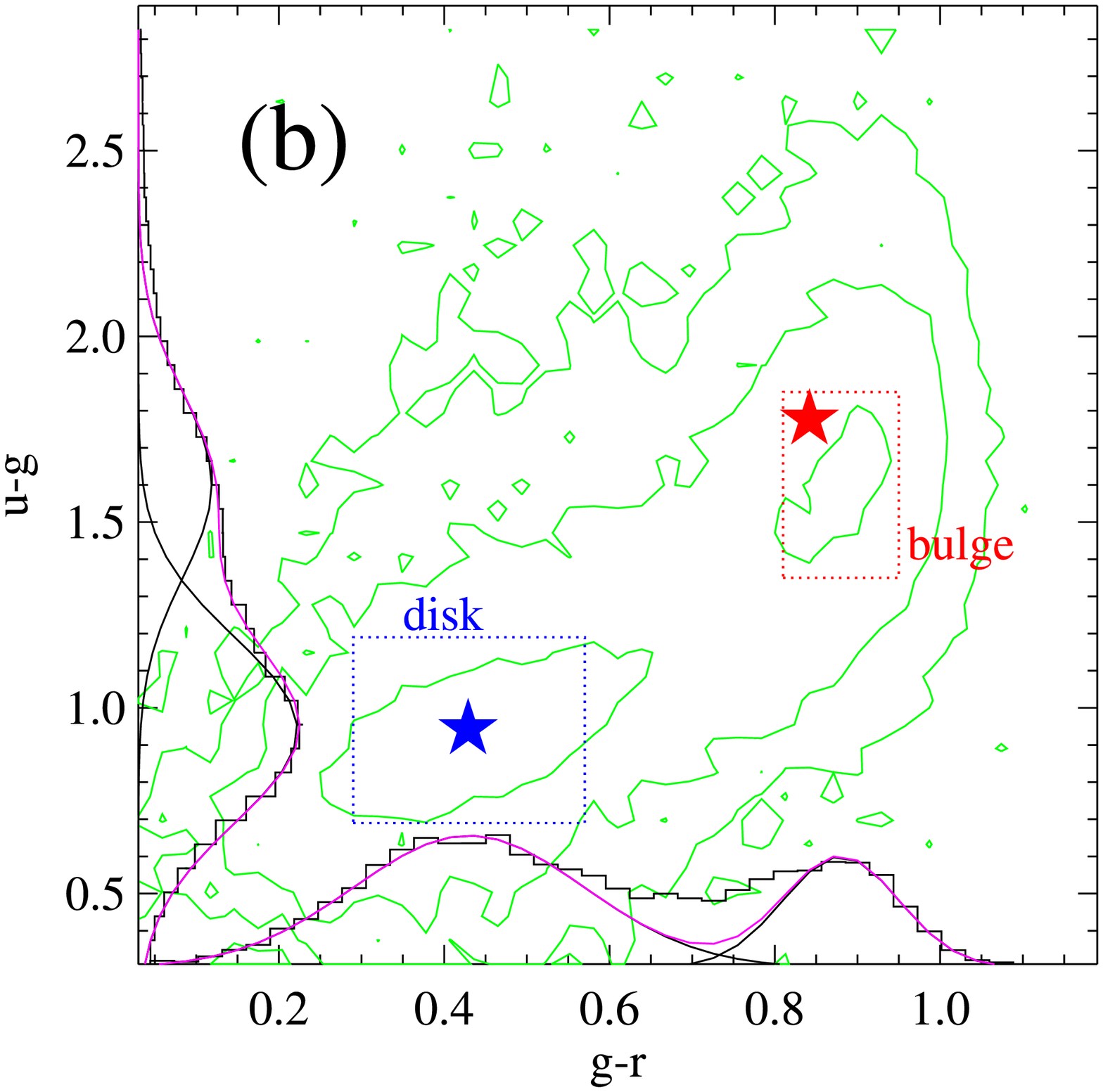}}%
\caption{
{\footnotesize
{\it Panel (a):} 
Color-color diagram in $\ug$ vs. $\gr$ plane. 
The three lines represent the delayed exponential 
decay models with Chabrier IMF for metallicities: 
$Z/Z_\odot = 0.2$ ({\it black open squares}),  
1.0 ({\it blue solid triangles}), 
and 2.5 ({\it red open triangles}), with $\tau$ 
running along the line from top to bottom as indicated in the legend.  
The contour shows the color distribution for SDSS galaxies 
taken from \citet{Blanton03b}.
{\it Panel (b):} 
Same as (a), but with the histogram for the color 
distribution of SDSS galaxies projected onto the axes 
together with Gaussian fits. The two boxes indicate 
the location of the peaks of the histogram, 
with the widths twice the dispersions of the Gaussian. 
The asterisks indicate the bulge component 
(with $\tau=1.5$\,Gyr, $Z/Z_\odot=1.5$) and the disk component 
(with $\tau=4.5$\,Gyr, $Z/Z_\odot=0.8$) of the `Fossil' model.
}}
\label{fig:colcol}
\end{center}
\end{figure}

We then compute the colors of galaxies for each model using BClib03
with the Chabrier IMF.
The delayed exponential models give $\ug$ and $\gr$ colors as shown 
in Figure~\ref{fig:colcol}a at $t_0=13.5$\,Gyr.
The three lines stand for metallicities: 
$Z/Z_\odot = 0.2$ ({\it black open squares}),  
1.0 ({\it blue solid triangles}), and 
2.5 ({\it red open triangles}), with $\tau$ varying 
along the line from top (0.1 Gyr) to bottom (10 Gyr) 
as indicated in the legend.  
Also shown as contours is the color distribution 
of SDSS galaxies taken from Figure~7 of \citet{Blanton03b}. 

The histogram (normalized arbitrarily) of the SDSS galaxy color 
projected onto each axis is shown in Figure~\ref{fig:colcol}b, 
with double Gaussian fits by eye.
The two rectangular boxes delineated with dots indicate 
the locations of the Gaussian peaks with the width 
twice the dispersion: 
$\ug = 0.94 \pm 0.25 $ and $\gr = 0.43 \pm 0.14$ for the blue peak,
and $\ug=1.6 \pm 0.25$ and  $\gr = 0.88 \pm 0.07 $ for the red peak. 
Much has been said of the bimodal color distribution seen in the 
local galaxies from the SDSS  \citep{Strateva01, Blanton03b, 
Kau03b, Baldry04, Balogh04, Brinchmann04}. 
The two distinct distributions correspond to non-star forming 
and star-forming galaxies, but roughly also to the redder 
spheroid-dominated component and the bluer disk galaxies. 
This is the concept first noticed by \citet{Baade44} 
in his proposal of the two population model.  
He also pointed out that the two types of stellar populations 
had been recognized as early as 1926 by \citet{Oort26}. 

We represent the peaks of SDSS galaxy color distribution 
with two parameters $(\tau, Z/\Zsun)$. 
Figure~\ref{fig:colcol}a shows that the models with 
$\tau = 1.5$\,Gyr and 4.5\,Gyr fit the colors of the 
spheroid and disk components reasonably well, respectively, 
and the variation of metallicity gives a further tuning. 
We find that it is difficult to realize the colors 
$(\ug, \gr) = (1.6, 0.9)$ with BClib03 even when we push 
the metallicity to $Z/\Zsun=2.5$, which is unlikely to be 
a representative of spheroid component even if
the most massive elliptical galaxies may have such a high metallicity.  
We present the stellar mass weighted mean metallicity for spheroids
(early type galaxies and bulges of disk galaxies) of different 
morphological types in Table~\ref{table:bulge}, as
calculated in Appendix~\ref{sec:bulge_metal}.
As a result, we construct a composite `Fossil' model 
--- we call it Fossil because it is based on the fossil evidence 
in the present universe, i.e., colors of galaxies --- 
with the parameters ($\tau\,[\Gyr], Z/\Zsun) = (1.5, 1.5)$
for the spheroid component and $(4.5, 0.8)$ for the disk component
\footnote{Here we are representing the entire galaxy population 
with two delta functions on the color-color plane. 
It is possible in the future to extend Figure~\ref{fig:colcol}
into a three dimensional color space by adding another color
as a z-axis, and assign a set of parameters ($\tau, Z$) 
to every grid cell in the 3-d color space. That would be a 
more smooth, continuous representation of stellar mass 
distribution in the color space.}. 
These parameters are indicated by the {\it asterisks} 
in Figure~\ref{fig:colcol}b.  The blue asterisk for the 
disk component is located at the center of the box, but 
the red asterisk for the spheroid is slightly off to the corner. 
The mass-weighted mean metallicity of the spheroid component
we obtain in Table~\ref{table:bulge} is $1.3\,\Zsun$, but
in order to keep the red asterisk within the dotted box
in Figure~\ref{fig:colcol}b, we choose to adopt 
$1.5\,\Zsun$ for the spheroid component. 

Next, we determine the normalization of the two components.
We choose to normalize each component by requiring a match 
of the energy output in $K$- and $r$-bands with observations 
(see Appendix~\ref{sec:fossil_norm}). 
Our normalization corresponds to the bulge(spheroid)-to-disk
stellar mass ratio $(M_b/M_d)=1.35$, or equivalently 
$M_b:M_d=57\%:43\%$.  An alternative method to determine
the normalization uses the bulge-to-disk luminosity ratio 
and the mass-to-light ratio for each morphological type of
galaxies, as outlined in the caption of 
Table~\ref{table:bulgetodisk}; 
this gives $(M_b/M_d)=1.9$, or $M_b:M_d=66\%:34\%$.
Given the uncertainties involved, we consider 
the bulge stellar mass fraction of $55-70$\% to be
observationally allowed. 
For this work we adopt 57\% obtained in 
Appendix~\ref{sec:fossil_norm}, but the exact value 
is not important to the major conclusions.

\begin{figure}
\epsscale{1.0}
\plotone{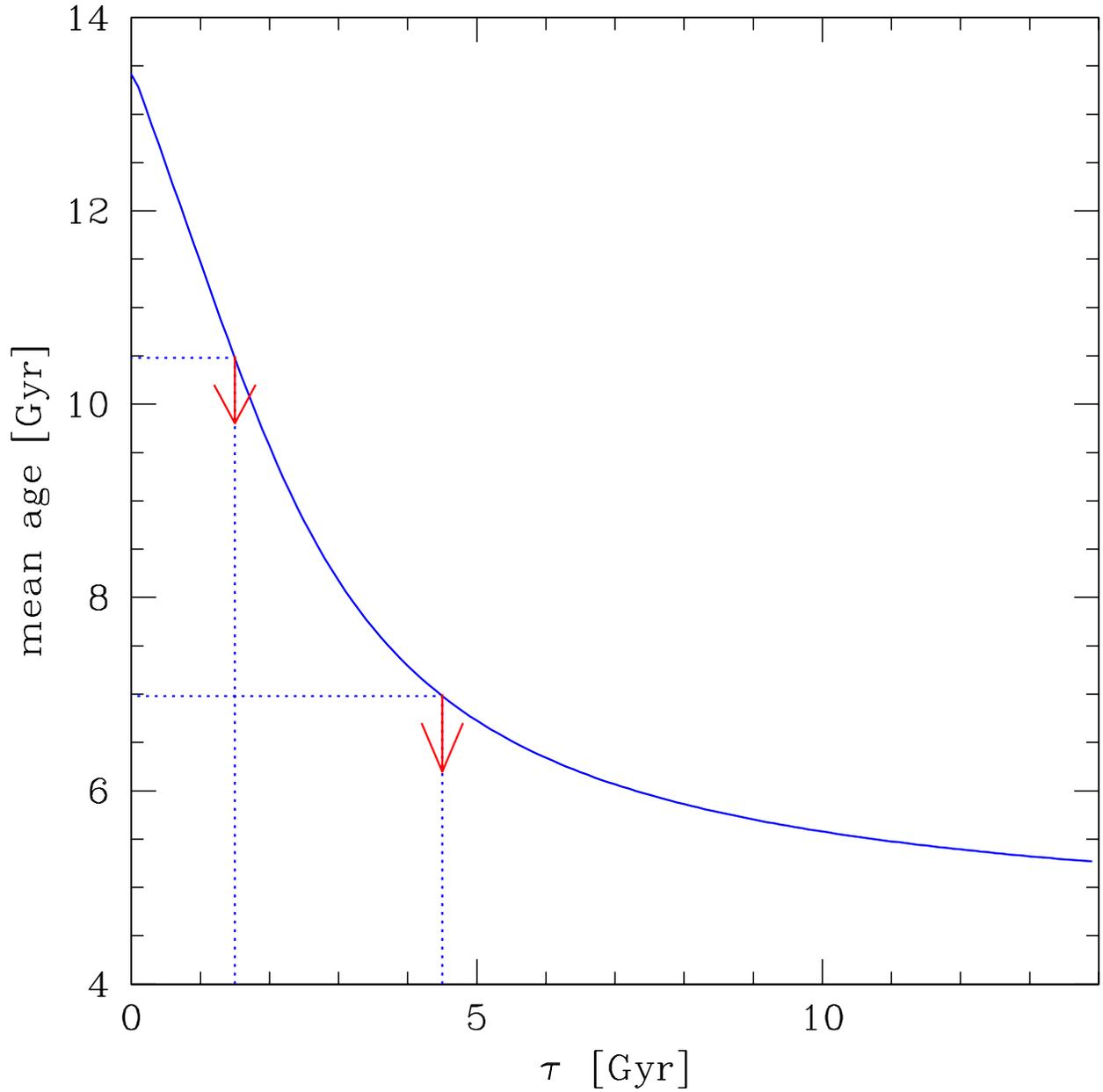}
\caption{
Mean age of stars as a function of 
decay time $\tau$ for the delayed exponential 
model. The dotted lines shows the values of $\tau$ 
($\tau = 1.5$\,Gyr and 4.5\,Gyr) 
adopted for the `Fossil' model. 
The arrows indicate the shift in the mean age 
when stellar remnants are removed from the calculation, 
while the curve includes the contribution from 
the dead stars that are not emitting light 
at the present time, therefore push the age up.
} 
\label{fig:age}
\end{figure}

Figure~\ref{fig:age} shows the mass-weighted mean stellar age 
for the delayed exponential model, calculated as 
\begin{equation}
t_{\rm age} = t_0 - \left (\int_0^{t_0} t\,\dot{\rhos}(t)\,dt\ / 
\int_0^{t_0} \dot{\rhos}(t)\,dt \right ). 
\end{equation}
The mean age of disk stars ($\tau=4.5$\,Gyrs) turns out to be 
$t_{\rm age}=7$\,Gyrs, which is consistent with the 
observational estimate of $5-7$ Gyrs \citep{Rocha00, Robin03, Naab06a}.
We note that our estimate includes the stellar remnants. 
Had one omitted the contribution from stars that died 
in the history of the universe, then the mean age 
would be younger by approximately 0.8\,Gyr, as indicated by 
the arrows. The main properties of our Fossil model 
at the present epoch are summarized in Tables~\ref{table:fossil} 
and \ref{table:params}. 
It may be helpful to note that the historical development of the 
present distribution of galaxies is of no import for the Fossil model. 
It makes no difference at all to the accounting if we had dry, wet, 
or no mergers. All that matters is that the local SDSS inventory
of a representative volume enables us to estimate the age distribution
of star formation.


\section{Cosmic star formation history}
\label{sec:sfr}

\begin{figure}
\epsscale{1.0}
\plotone{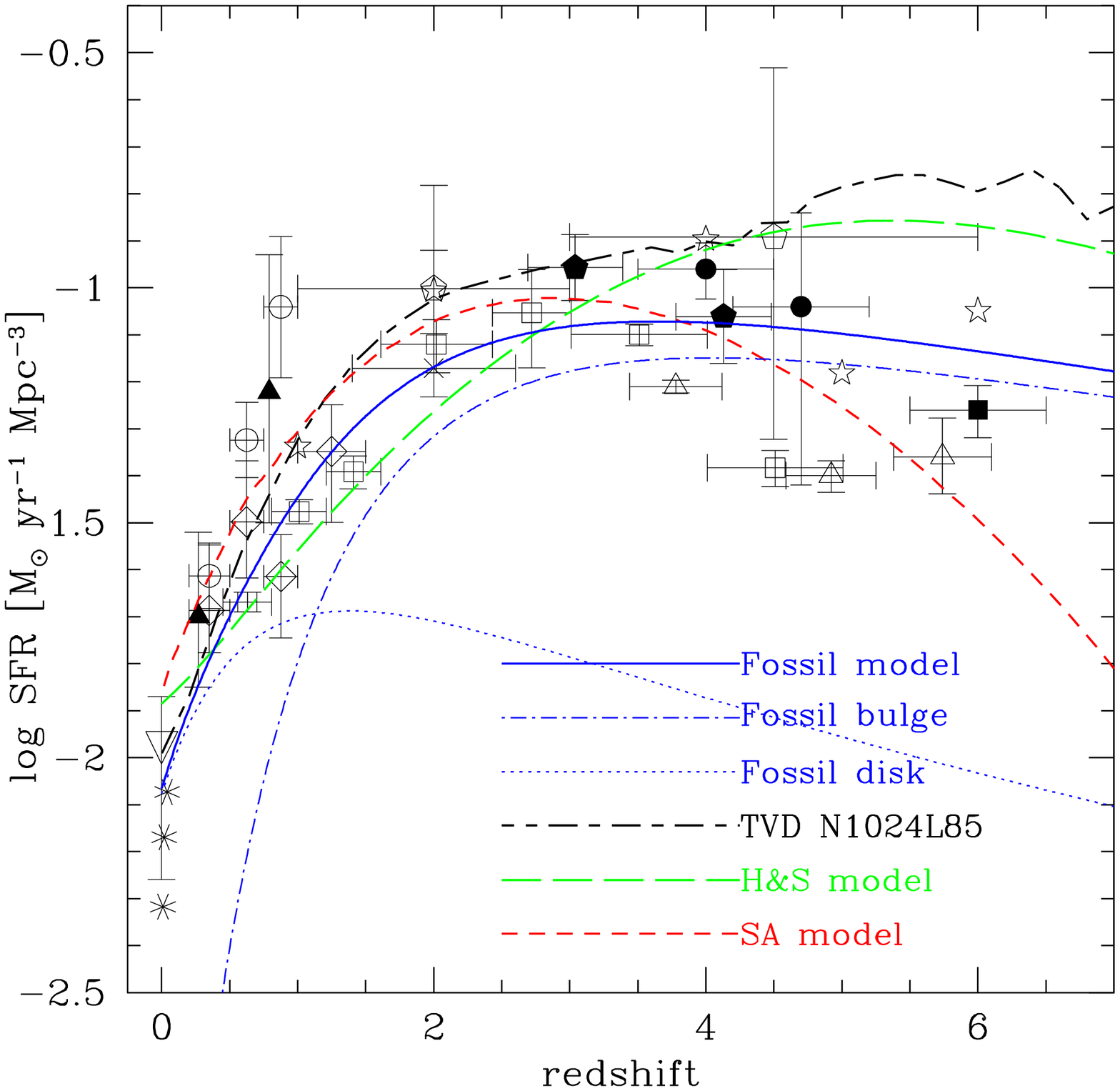}
\caption{
Star formation rate density as a function of redshift.
The curves represent the model predictions specified in the legend.
The data are taken from (from low to high redshift): 
\citet[][3 asterisks at $z\sim 0$]{Heavens04}, 
\citet[][open inverted triangle at $z=0$]{Nakamura04},
\citet[][open circles]{Lilly96}, 
\citet[][filled triangles]{Norman04},
\citet[][open diamonds]{Cowie99},
\citet[][open squares]{Gabasch04},
\citet[][cross at $z=2$]{Reddy05}, 
\citet[][open pentagons at $z=2$ and 4.5]{Barger00}, 
\citet[][filled pentagons at $z=3,4$]{Steidel99}, 
\citet[][filled circles at $z=4,5$]{Ouchi04a}, 
\citet[][open triangles at $z=3-6$]{Giavalisco04}, 
\citet[][filled square at $z=6$]{Bouwens05}, 
and 
\citet[][open stars without error bars]{Thompson06}.  
The data are converted to the values with the Chabrier IMF and 
common values are assumed for dust extinction for the UV data.  
See text for details. 
}
\label{fig:sfr}
\end{figure}

In Figure~\ref{fig:sfr}, we show the cosmic SFR density as a function 
of redshift. 
The {\it blue solid} line is the Fossil model described in the previous
section, and the {\it black short-dash long-dash} line shows the result of 
a new Eulerian TVD hydrodynamic simulation. This simulation has a 
comoving box size of $85\himpc$ and $1024^3$ hydrodynamic mesh, 
and the code is similar to the one used by \citet*{Cen05} and 
\citet{Nachos1}. The physical cell size is $83/(1+z)\,\hikpc$. 
The mean baryonic mass per cell for this simulation 
is $m_{\rm gas}= 3.63\times 10^6\, \himsun$ and the dark matter particle 
mass is $m_{\rm DM}= 1.58\times 10^8\, \himsun$ ($512^3$ particles). 
The cosmological parameters are 
$(\Om, \Ol, \Ob, h, n, \sigma_8)= 
(0.31, 0.69, 0.048, 0.69, 0.97, 0.89)$.
The TVD result is extracted from the simulation without additional 
processing except for the boxcar smoothing over $\pm 3$ bins in the
redshift axis. 
The amount of gas converted into stars is a direct output of the 
simulation, so there is no uncertainty as to stellar IMF
and no freedom for the normalization.
For the details on the star formation and supernova (SN) feedback
prescription, we refer the readers to \citet{CO93, Nag01a, Cen05}. 

We also consider two other models: one from 
\citet[][H\&S model]{Her03} who derived an analytic approximation 
for the SFR history as a function of redshift based on their 
cosmological Smoothed Particle Hydrodynamics (SPH) simulations.
The H\&S model shown in {\it green long-dashed} line takes the form 
\begin{equation}
\dot\rhos = \dot\rhos(0)
\frac{\chi^2}{1+\alpha(\chi-1)^3\exp{(\beta\chi^{7/4})}},
\label{eq:sfr}
\end{equation}
where $\chi(z) \equiv (H(z)/H_0)^{2/3}$. 
With star formation and feedback models adopted in 
\citet{SH03a}, the parameters are $\alpha=0.012$, $\beta=0.041$, and
$\dot\rhos(0)=0.013\,{\rm M_\odot ~yr^{-1}~Mpc^{-3}}$.
The normalization is fixed to give the local SFR density \citep{Her03}.

The other model represented by {\it red short-dashed} line is the 
semi-analytic model GALFORM by \citet[][hereafter SA model]{Cole00}. 
The cosmological parameters that H\&S and SA adopted
$(\Om, \Ol, \Ob, h, n, \sigma_8)=
(0.3, 0.7, 0.04, 0.7, 1.0, 0.9)$,
differ slightly from those of the TVD simulation.  
We confirmed that this slight difference is not important for 
our analyses. 
Table~\ref{table:params} summarizes the basic characteristics of 
these models.

For the observational data shown in Figure~\ref{fig:sfr}, we assume 
the Chabrier IMF ($0.01-100\Msun$).
The UV luminosity densities $\rhouv$ are converted into SFR by 
\begin{equation}
\rhouv\ [\erg\ \s^{-1}\, \Hz^{-1} \mpc^{-3}] = 
C\, \dot{\rhos}\ [\Msun\, \yr^{-1}\, \mpc^{-3}],
\end{equation} 
where the parameter $C$ can be derived from BClib03 as
\begin{equation}
C = 1.24 \times 10^{28}
\label{eq:sfrconversion}
\end{equation} 
for an exponentially decaying star formation history with
$\tau=5$\,Gyr for the Chabrier IMF and solar metallicity. 
Here $\rhouv$ is computed by averaging the flux over 
$\pm 150$\,\AA\ centered at $\lam = 1500$\,\AA\ at age $t=10$\,Gyr.
The parameter $C$ depends on the age of the stellar population
and reaches a plateau after $t\sim 100$\,Myr
for $\tau \gtrsim 1$\,Gyr models. 
The value in equation~(\ref{eq:sfrconversion}) differs 
by a factor 1.6 from the value given by \citet*{Madau98} 
who performed the same calculation for the Salpeter IMF 
($0.1 - 100\,\Msun$).
 
The correction due to dust obscuration is important. 
For observations of nearby universe, the Balmer emission lines 
\citep{Gallego95, Tresse98, Hopkins00, Pascual01, Tresse02, 
Nakamura04} can be used to estimate the SFR with 
reasonable reliability at $z\sim 0$.  
For higher redshift, where SFR is derived from the UV light,
estimates of extinction are more uncertain. 
\citet{Steidel99} took $E(B-V)=0.15$, somewhat smaller than 
the value inferred for the local galaxies.  
At some point at high redshift, the dust content should
start decreasing as a function of increasing redshift, 
but whether this is seen towards higher redshift or not 
is controversial even today.  
\citet{Bouwens05} suggest a modest decline in the dust extinction effect 
from $z=3$ to $z=6$, but \citet{Thompson06} indicate 
the extinction correction of a factor of 5 at $z=6$, 
which differs little from that at $z=3-4$. 
In this situation we must be content with with our fiducial choice 
of the conventional prescription of \citet{Steidel99} 
that the SFR are corrected by factors of 2.7 ($z<2$) and 4.7 ($z>2$), 
as also supported by the subsequent work by \citet{Reddy04}. 
We apply this correction to most of the data
(after removing the corrections by individual authors), 
keeping in mind that varying dust extinction 
as a smooth function of redshift is
needed to obtain more reliable estimates of SFR.
Note that the faint-end slope of the luminosity 
function at high-redshift is poorly constrained 
and SFRs are derived under different assumptions 
(we do not dare to standardize them except for extinction 
corrections), which we review briefly in what follows.

\citet[][filled pentagons at $z=3, 4$]{Steidel99} derived their UV 
luminosity density by integrating the luminosity function 
with the faint-end slope of $\alpha=-1.6$ down to $0.1 L^*$. 
We take the SFR from their Fig.9, and convert it to the 
$\Lam$ cosmology. 
\citet[][open triangles at $z=3-6$]{Giavalisco04} 
integrated the Schechter fit with $\alpha=-1.6$ to $0.2 L^*$. 
\citet[][filled circles]{Ouchi04a} integrated 
the luminosity function with $\alpha=-2.2$ to $0.1 L^*$. 
\citet[][filled square at $z=6$]{Bouwens05} integrated their 
Schechter fit with $\alpha=-1.74$ to $0.04 L^*_{z=3}$. 
Their sample goes to the faintest magnitude compared to others, 
yet the result still indicates some decline in SFR density 
from $z=3$ to $z=6$ by a factor of $\sim 0.7$.  
\citet{Thompson06} performed the SED fitting to 
the 6 photometric broadband measurements by ACS and NICMOS 
of the Hubble Ultra Deep Field determining the extinction 
for each sources.  They then utilized the star formation 
intensity distribution function \citep{Lanzetta02} to correct
for the surface brightness dimming effect.  Their result is
consistent with the earlier result by \citet{Thompson03}. 
Here we take the extinction uncorrected data listed 
in their Table 2 and applied our correction as mentioned
above.  Since their errors are large and the error bars 
overshoot to the outside of the plotted range of 
Figure~\ref{fig:sfr} when put on our version of data points, 
here we omitted the error bars for this data.

Extinction corrections are not applied to the data derived from 
X-ray \citep{Norman04} and submillimeter observations 
\citet[][taken from Table~2 of \citet{Hopkins04}]{Barger00},
as they are less subject to dust extinction effects. 
The data by \citet{Heavens04} at very low-redshift from the MOPED 
\citep{Heavens00} algorithm already include the correction for 
dust for each galaxy.
The data of \citet{Nakamura04} are derived from hydrogen Balmer lines 
including a consistent correction for extinction in individual
galaxies.
We note that uncertainties for $z\geq 1$ could be much larger 
than those represented here, while those for lower redshift 
are probably not too large since dust extinction is constrained well 
with the use of Balmer lines.

With large uncertainties we discussed in mind,
Figure~\ref{fig:sfr} shows a good agreement among the three 
independent approaches: observed SFR, Fossil model, and 
{\it ab initio} models. 
All estimates agree with each other to within about a factor of two,
which is also the size of the scatter in the data.  
We see some declining trend of the SFR in the SA model at $z>4$,
which contrasts to the other numerical {\it ab initio} models 
and the Fossil model that show a roughly constant rate of star formation. 
Our argument given here should be understood in the sense that 
if observations would receive too large systematic errors, 
the consistency with the two other approaches is endangered.

We also underline the conclusion from the Fossil model that the
spheroidal component formed primarily at high-redshift ($z \gtrsim 1.5$), 
and the majority of the disk stars formed at low-redshift ($z \ltsim 1$). 
The sum of these two components falls in-between the curves of other models.

A note on the uncertainty due to our poor understanding of star 
formation is appropriate. The theoretical calculations are most 
secure in their predictions of the SFR, while the prediction of 
the electromagnetic output (and the stellar mass density to some 
extent) do depend on assumptions concerning the 
stellar IMF.   The observational measurements are done for the 
electromagnetic output and SFR is derived by assuming
an IMF, therefore the observational estimates of SFR 
suffers from the uncertainty in the assumed IMF. Further
uncertainty is expected from dust extinction. This is
particularly important when we deal with far UV light.

We also remark that the star formation rate in the simulations
could depend on the details of the SN feedback prescriptions. 
If the feedback is turned off in the simulations, 
the heating of the gas is underestimated and therefore the
global SFR density at low redshift is overestimated 
(i.e., {\it physical} overcooling of the gas). 
In the TVD simulation used in this paper and the SPH simulations
on which the H\&S model is based, the strength of the feedback
was set based on the energetic argument of SN explosion 
and was also calibrated against various observations 
such as the star formation rate of Lyman break galaxies,  
metallicities of the Ly$\alpha$ forest and intra-cluster gas, 
and the galactic wind speeds observed in starburst galaxies. 
The current spatial resolution of the TVD simulation 
(physical cell size of $83/(1+z)\,\hikpc$) is 
not adequate to resolve the details of the inner structure of 
galaxies, but the overall star formation rate can be simulated
with some confidence as the energetics are faithfully simulated
on scales slightly larger than the real galaxies.


\section{Stellar mass density} 
\label{sec:omegastar}

\begin{figure}
\begin{center}
\resizebox{9.5cm}{!}{\includegraphics{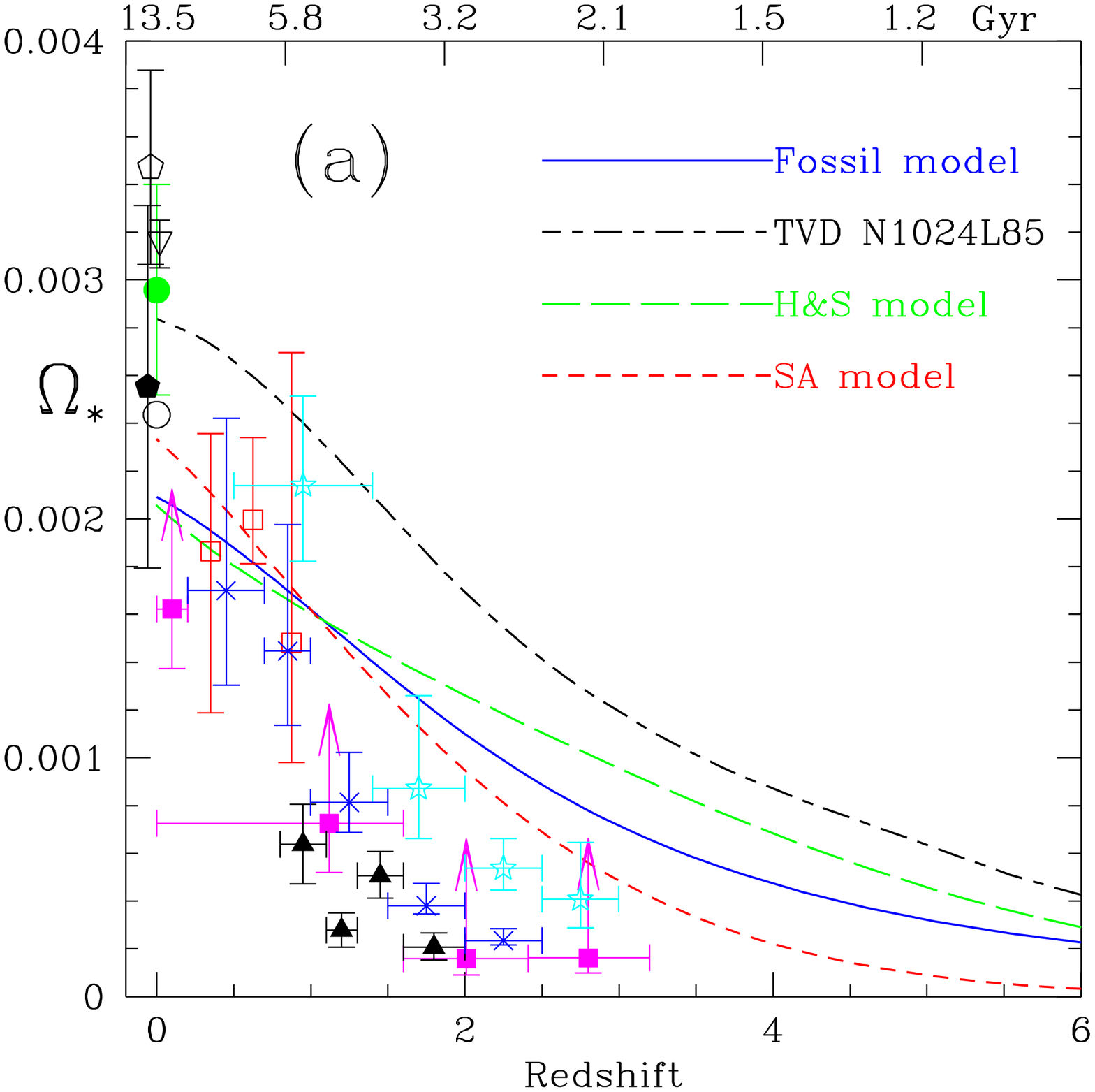}}\\
\vspace{0.2cm}
\resizebox{9.5cm}{!}{\includegraphics{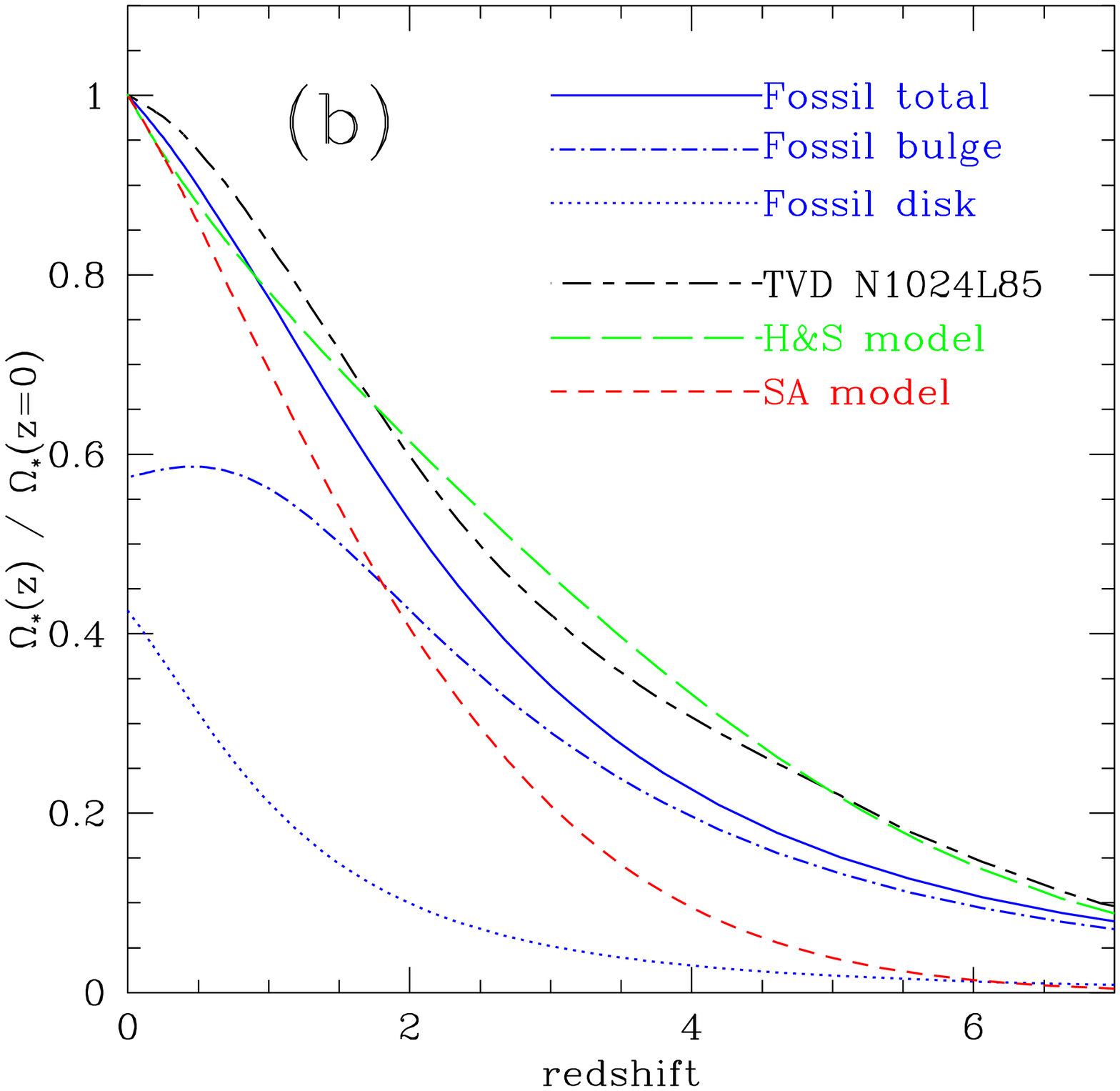}}%
\vspace{0.1cm}
\caption{{\it Panel (a):} Stellar mass density (including 
remnants) as a function of redshift and age. 
The top axis indicates the age of the Universe.
All data assume the Chabrier IMF and $h=0.7$.
The source of data are given in the text. 
{\it Panel (b):} Growth of the stellar mass density 
normalized by the value at $z=0$. 
For the Fossil model, the bulge and disk components
are also shown separately.
}
\label{fig:omega}
\end{center}
\end{figure}

As we noted, the integral of SFR from $t=0$ to $t_0$ is not 
equal to the stellar mass density, as the gas recycles
into the interstellar medium. 
Figure~\ref{fig:omega}a shows the evolution of the global stellar mass 
density $\Os$ as a function of redshift. 
The model results take account of the time-varying recycling fraction 
derived from BClib03. All results assume the Chabrier IMF 
($0.01 - 100\,\Msun$).

The values of $\Os$ at $z=0$ after the gas recycling correction are 
0.0021, 0.0028, 0.0021, and 0.0023 for the Fossil, TVD, H\&S, and 
SA models, respectively (see Table~\ref{table:params}).
The empirical estimates converges to 
$\Os \approx (2.4 - 3.6)\times 10^{-3}$ with 
the Chabrier IMF ($0.01-100\,\Msun$), or 
$(1.8 - 3.9)\times 10^{-3}$ taking the errors into account.
The model values are consistent with this range
with a lower value favored. 
We remark that the conversion factor for $\Os$ 
from Salpeter IMF (mass range $0.1-100\,M_\odot$) to 
Chabrier IMF ($0.01-100\,\Msun$) is 1.4
for the same amount of luminosity output at $M>1 \Msun$. 
The data are taken from
\citet[][{\it green filled circle}]{Cole01}, 
\citet[][{\it black open circle}, the size of the claimed error bar 
is comparable to the symbol size]{Panter04},
\citet[][{\it black inverted triangle}]{Fukugita04},  
\citet[][{\it black open pentagon}, 
slightly offset from $z=0$ for clarity]{Kochanek01}, and 
 \citet[][{\it solid pentagon}]{Bell03}.

The situation is different for the stellar mass for higher $z$. 
The figure compares the model curves with the observational analyses 
given by  
\citet[][{\it magenta filled squares}, Table~2]{Rudnick03}, 
\citet[][{\it cyan open stars}]{Dickinson03b}
\citet[][{\it red open squares}]{Brinchmann00}, 
\citet[][{\it blue open crosses}, Table~4, `observed' column]{Fontana04}, 
and
\citet[][{\it black filled triangles}]{Glazebrook04}
all transformed to the Chabrier IMF.

At $z\gtrsim 1$ the observational estimates all fall significantly short 
of the model. 
Roughly speaking, all models predict that $\sim 60$\% 
of the present stellar mass was formed by $z=1$,
whereas the observation indicates it is as small as  $20-30$\%, 
a gross disagreement, as also discussed by \citet{Nachos1} earlier.
We emphasize that this is likely to be
an observational problem, since the straightforward integration 
of the empirical
star formation rate, corrected for the recycling factor, yields a
stellar mass substantially larger than what is observed, indicating 
a gross underestimate of the stellar mass in high-redshift galaxies 
in currently available analyses.
We note that possible errors in extinction corrections 
for $z\geq 1$ galaxies do not solve this problem.


\section{Metal mass density}
\label{sec:metal}

\begin{figure}
\epsscale{1.0}
\plotone{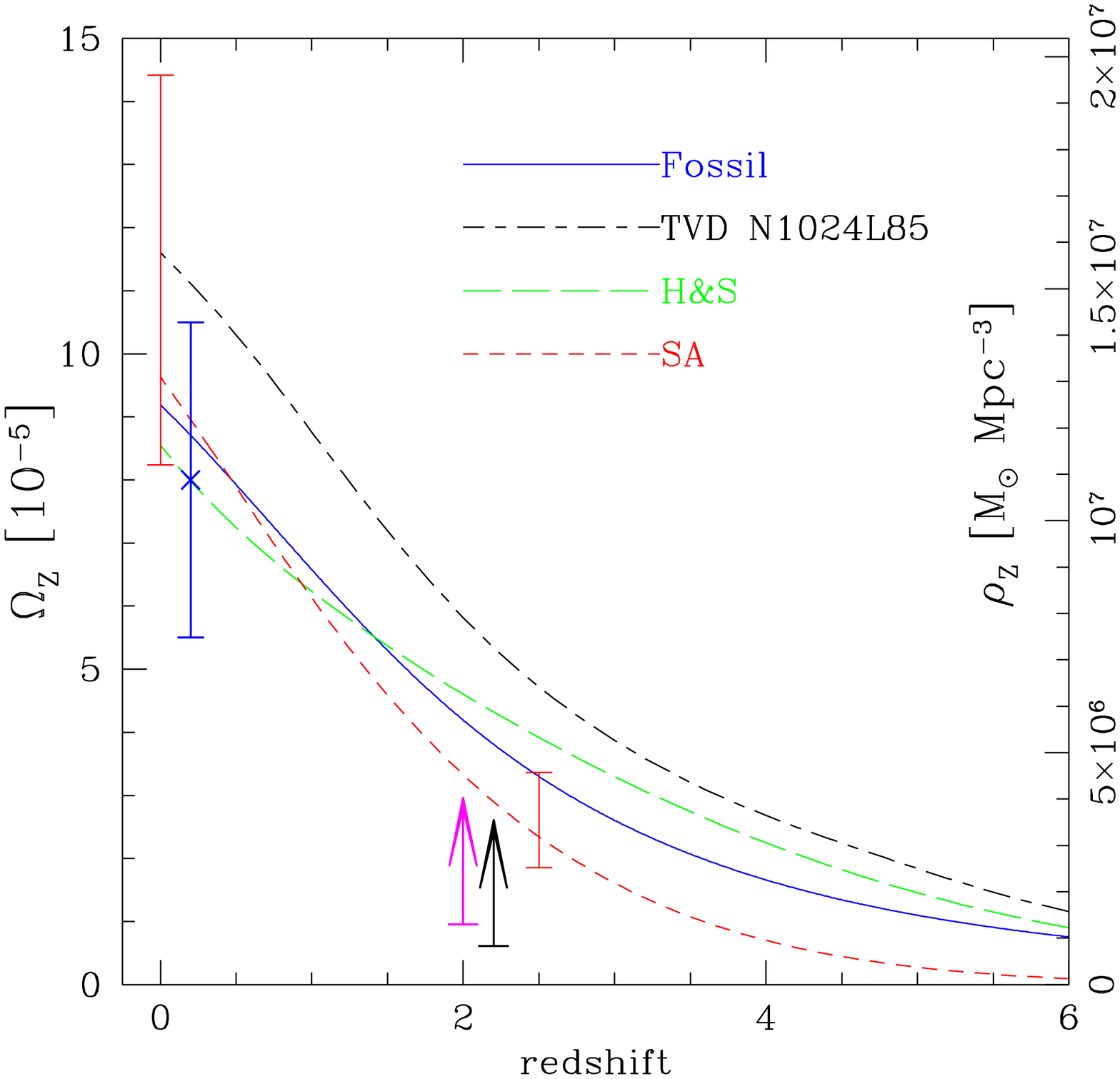}
\caption{
Metal mass density as a function of redshift.
The curves represent the model predictions specified in the legend.
The data are taken from (from low to high redshift): 
\citet[][red bars at $z=0$ \& 2.5]{Dunne03}, 
\citet[][blue cross at $z=0$, shifted for clarity]{Fukugita04},
\citet[][magenta lower limit at $z=2$, for galaxies]{Bouche06}, 
and 
\citet[][black lower limit at $z=2$, shifted for clarity. Only for 
damped Ly$\alpha$ systems and super Lyman Limit systems]{Pro06}. 
}
\label{fig:metal}
\end{figure}

Given the star formation histories, one can compute the metal mass
density as a function of redshift as
\begin{equation}
\dot{\rhos} = Y \dot{\rhoz},
\end{equation}
where we take $Y=0.023$ using the prescription of 
\citet[][Section 14.4]{Arnett} adjusted to the Chabrier IMF. 
(This value corresponds to the Searle-Sargent yield $Y/(1-R)=0.042$.) 
The evolution of metal mass density is shown in Figure~\ref{fig:metal}
together with empirical estimates.

At $z=0$, all models agree well with the estimate by 
\citet{Fukugita04}, 
$\Oz=(0.8 \pm 0.25) \times 10^{-4}$ (eq.[94]), 
which excludes the contributions from stellar remnants
but includes those from warm galactic halos.
\citet[][Table~1]{Dunne03} give an estimate higher than 
\citet{Fukugita04} by 50\%, and the TVD model is consistent
with their central value. 
Other model predictions are consistent with 
Dunne et al.'s estimate at its lower edge.

\citet{Dunne03} also give an estimate of the metal abundance 
at $z=2.5$ including damped Ly$\alpha$ systems.
The models are consistent with their estimate at $z=2.5$, 
although the TVD and H\&S model lie somewhat above 
the given error bar.
\citet{Bouche06} and \citet{Pro06} presented metal abundance 
from high-z galaxies and damped Ly$\alpha$ systems 
lower than Dunne et al. 
They are, however, taken as lower limit, first because of 
missing contributions of faint galaxies as discussed by 
the authors, but perhaps more importantly by the omission of 
metals in warm galactic halos, which may contribute 
by additional 50-100\%.  Therefore we consider 
that the "missing metal problem" is not really a problem.


\section{Energy in the radiation}
\label{sec:ebl}

The last test concerns the consistency with the energy in the
radiation field, the product of stellar evolution.
To estimate EBL we use the compilation of \citet{Hauser01} 
for the source of observations. We write the integral of the flux
over the range between $\lam_1$, and $\lam_2$ (in $\mu$m), as
\begin{equation}
\EBL[\lam_1,\lam_2] \equiv \int_{\lam_1}^{\lam_2} \nu I_{\nu}\, d\ln \nu
\end{equation}
in units of nW m$^{-2}$ sr$^{-1}$.
For the optical to the near IR region direct observations of extragalactic
background yield $\EBL[0.16,3.5]\approx 60$, whereas the integration of
galaxy counts give  $\EBL[0.16,3.5]\approx 18$. We may take these two 
numbers as upper and lower limits, since the former is derived from difficult 
observations that are apt to be contaminated with local emission, 
while the latter would miss light from the outskirts of galaxies.
For the far IR that is dominated by dust
we adopt Hauser-Dwek's estimate:
$\EBL[3.5,140] = 11-58$,
where the upper value comes from fluctuation measurements and the lower
from integration of resolved sources.
Measurements are more secure for the submillimeter region:
 $\EBL[140,1000]=15\pm 2$.
Adding the three together, we take the total obscured EBL flux to be
\begin{equation}
\EBL[0.16,1000] = 42-135~ {\rm nW\, m^{-2}\, sr^{-1}},
\label{eq:obsebl}
\end{equation}
 which ranges over a factor of 3. 
We avoid quoting the central value, since
this range represents predominantly systematic uncertainties and 
what value should be taken as central is rather 
a matter of interpretation. 

The bolometric EBL and the comoving bolometric luminosity density 
at redshift $z$, $\jbol(z)$, are related by  
\begin{eqnarray}
\EBL &=& \left ( \frac{c}{4\pi} \right )\int_0^{\infty}\jbol(z)
\left |\frac{dt}{dz}\right | \frac{dz}{1+z} \\
&=& 9.63\times 10^{-8} h^{-1}\int_0^{\infty} 
\left [\frac{\jbol(z)}{L_{\odot}{\rm Mpc}^{-3}} \right ]
H_0 \left |\frac{dt}{dz}\right | \frac{dz}{1+z} 
\quad [\,{\rm nW}\,{\rm m}^{-2}\,{\rm sr}^{-1}\,],
\label{eq:ebl}
\end{eqnarray}
where
\begin{equation}
H_0 \left |\frac{dt}{dz}\right | = \frac{1}{(1+z)\sqrt{\Om (1+z)^3 + \Ol}}
\end{equation}
for a flat $\Lam$ universe,
and $\jbol(z)$ can be obtained by
\begin{equation}
\jbol(t) = \int_0^t \dot{\rho}_*(\tau) \left[L_{\rm bol}(t-\tau)/M\right] d\tau,
\label{eq:bolum}
\end{equation}
with $\dot{\rho}_*(\tau)$ the comoving SFR density
in units of $\Msun\,\yr^{-1}\,\mpc^{-3}$, 
and $L_{\rm bol}(t)/M$ $(\erg\,\Msun^{-1}$) the bolometric luminosity 
per mass of a stellar population with age $t$.


\subsection{Bolometric luminosity density}
\label{sec:bolum}

\begin{figure}
\begin{center}
\resizebox{9.5cm}{!}{\includegraphics{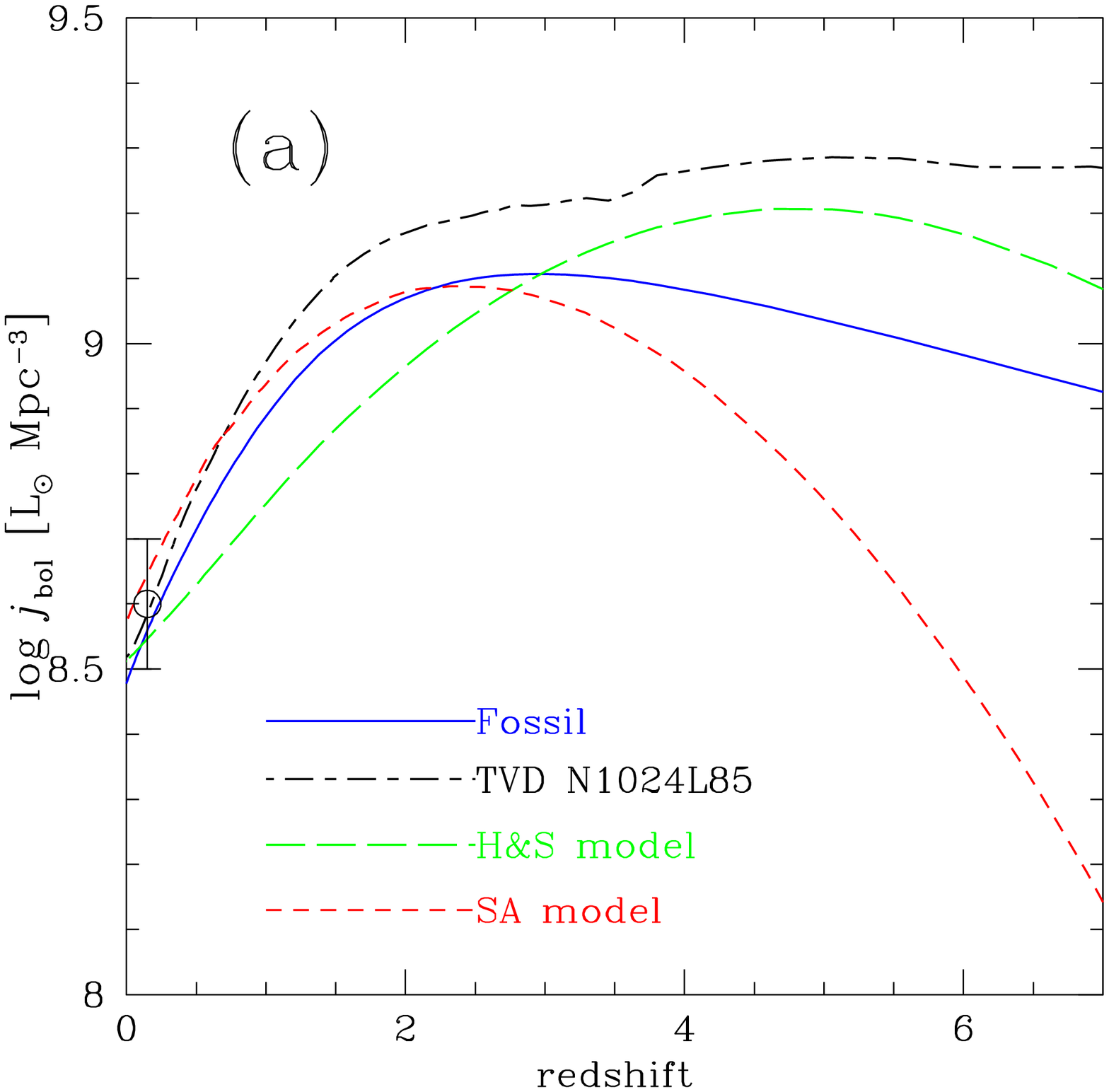}}%
\vspace{0.2cm}
\resizebox{9.5cm}{!}{\includegraphics{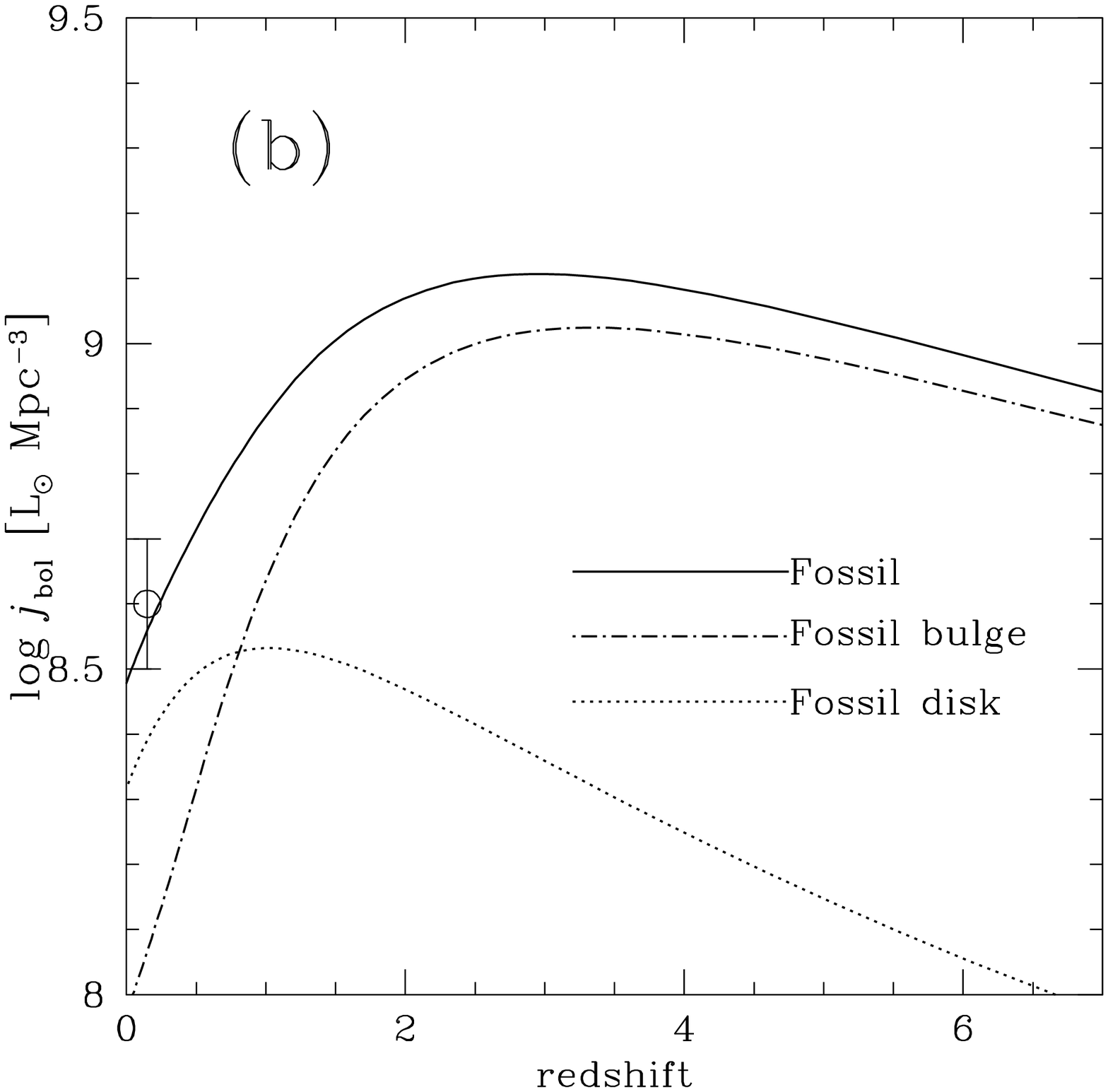}}%
\vspace{0.1cm}
\caption{{\it Panel (a):} 
Comoving bolometric luminosity density as a function of 
redshift for the models indicated in the legend.
{\it Panel (b)}: 
Same as (a) for the Fossil model, decomposed into the 
two components. 
The data at $z=0$ (slightly offset to a positive $z$ value 
for clarity) is the observational estimate given in 
Equation~(\ref{eq:jbol}) 
}
\label{fig:bolum}
\end{center}
\end{figure}

Figure~\ref{fig:bolum}a shows the comoving bolometric luminosity 
density as a function of redshift for the Fossil, TVD, H\&S, and 
SA models. For the latter three models a solar metallicity is assumed.
Figure~\ref{fig:bolum}b shows the spheroid and disk components 
separately for the Fossil model.
We use the bolometric electromagnetic output to minimize 
uncertainties due to corrections for dust obscuration.

We take \citet{Kashlinsky05a}'s summary for observed 
luminosity density at the present epoch:  
$\jbol(0.2-2\mu{\rm m})= (9.8\pm 1.2)\times 
10^{41}\, h\, \erg\, \s^{-1} \mpc^{-3}$ and $\jbol(12-100\mu{\rm m})= 
(1.5\pm 0.3)\times 10^{41}\, h\, \erg\, \s^{-1} \mpc^{-3}$. 
Adding these two and using $h=0.7$, we obtain 
$\jbol(0.2-100\mu{\rm m}) = (3.98\pm 0.43) \times 
10^8 L_{\odot, {\rm bol}} \mpc^{-3}$.  
Note that the energy density in the range of the gap, $2-12\mu$m is
expected to be small from the observations of the EBL.
We expect the contribution from $\lam > 100$ $\mu$m as
$\EBL[120, 1000] = 10-15$ nW m$^{-2}$ sr$^{-1}$ 
\citep{Hauser01, Kashlinsky05a}. 
This yields
$\jbol(0.2-1000\mu m) = (4.4\pm 0.5)\times 
10^8 L_{\odot, {\rm bol}} \mpc^{-3}$. 
Being conservative on the error, we adopt 
\begin{equation}
\log \jbol = 8.6\pm 0.1,
\label{eq:jbol}
\end{equation}
as our estimate of the bolometric luminosity density at $z=0$, 
which is shown in Figure~\ref{fig:bolum}. 
This estimate agrees with that of \citet[][see their Fig.15]{Bell03}. 
It is interesting that there is no discrepancy amongst the models 
and between the models and observations at $z=0$, 
while the model predictions are rather divergent
to one order of magnitude at high redshift.


\subsection{EBL}
\label{sec:eta}

 From the bolometric luminosity density shown in 
Figure~\ref{fig:bolum}, we compute the total EBL 
using equation~(\ref{eq:ebl}) as a function of redshift
for each model of cosmic star formation history. 
Our calculation does not include the contribution from AGN, 
which amounts to 7\%, as estimated in Nagamine et al. (2006, in preparation).
The model predictions are all convergent on $\EBL \approx 37-51$ 
nW m$^{-2}$ sr$^{-1}$ at $z=0$, which are to be 
compared with the observation, 
$\EBL[0.16,1000]=42-135$ nW m$^{-2}$ sr$^{-1}$ quoted above.
This indicates that the models are consistent with the observation  
only at its lower edge, suggesting that the current EBL observations
might still be contaminated from non-cosmological sources, whereas
the integration over resolved sources yields the value 
theoretically expected
\footnote{See Note Added in the end of this paper.}.
The model prediction would even undershoot the observations
if the Salpeter IMF were adopted.

\begin{figure}
\begin{center}
\plotone{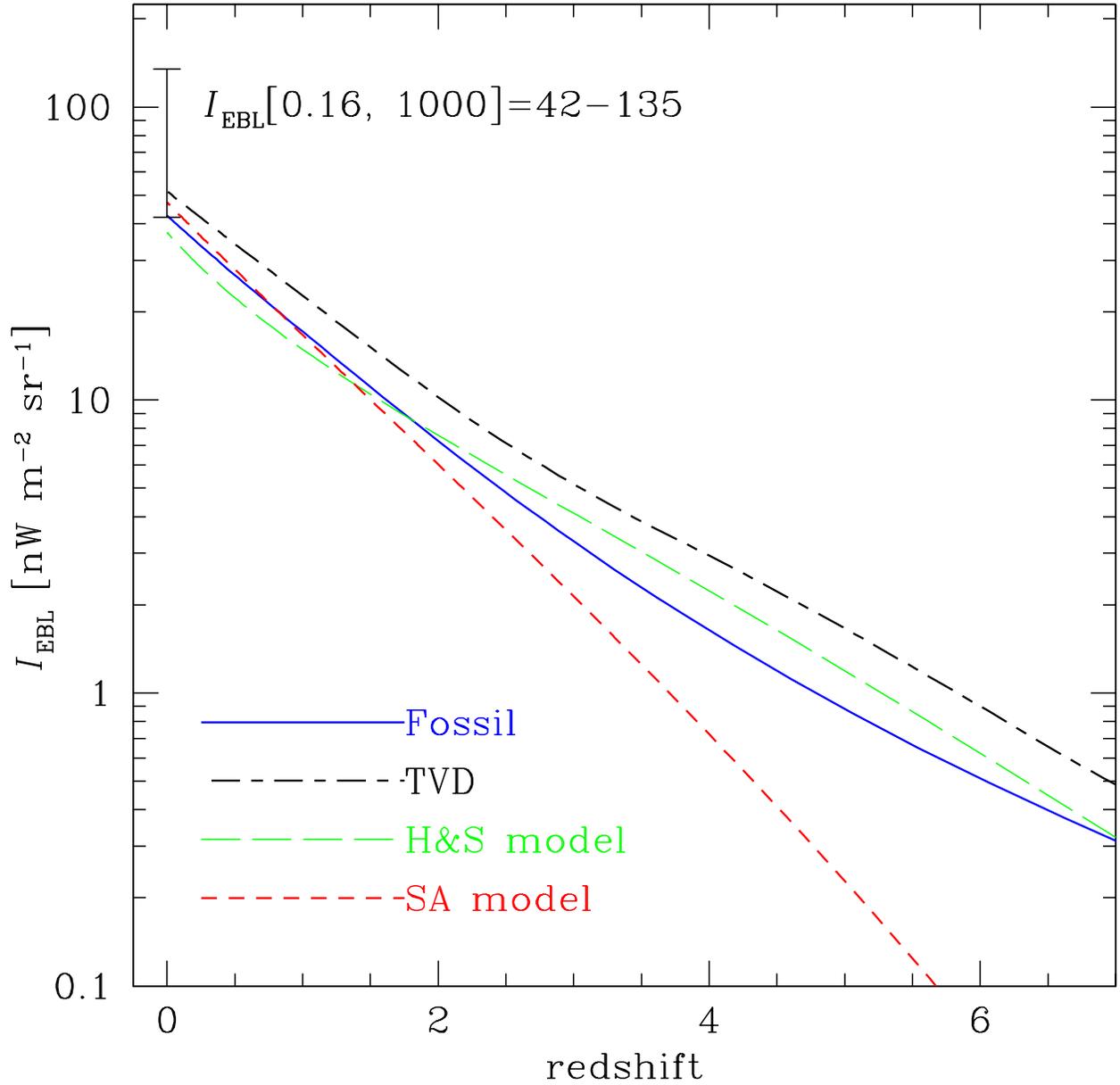}
\caption{EBL as a function of redshift 
for the models shown in the legend.
The observationally allowed range at $z=0$, 
equation~(\ref{eq:obsebl}), is indicated by the error bar.
}
\label{fig:ebl}
\end{center}
\end{figure}

\begin{figure}
\epsscale{1.0}
\plotone{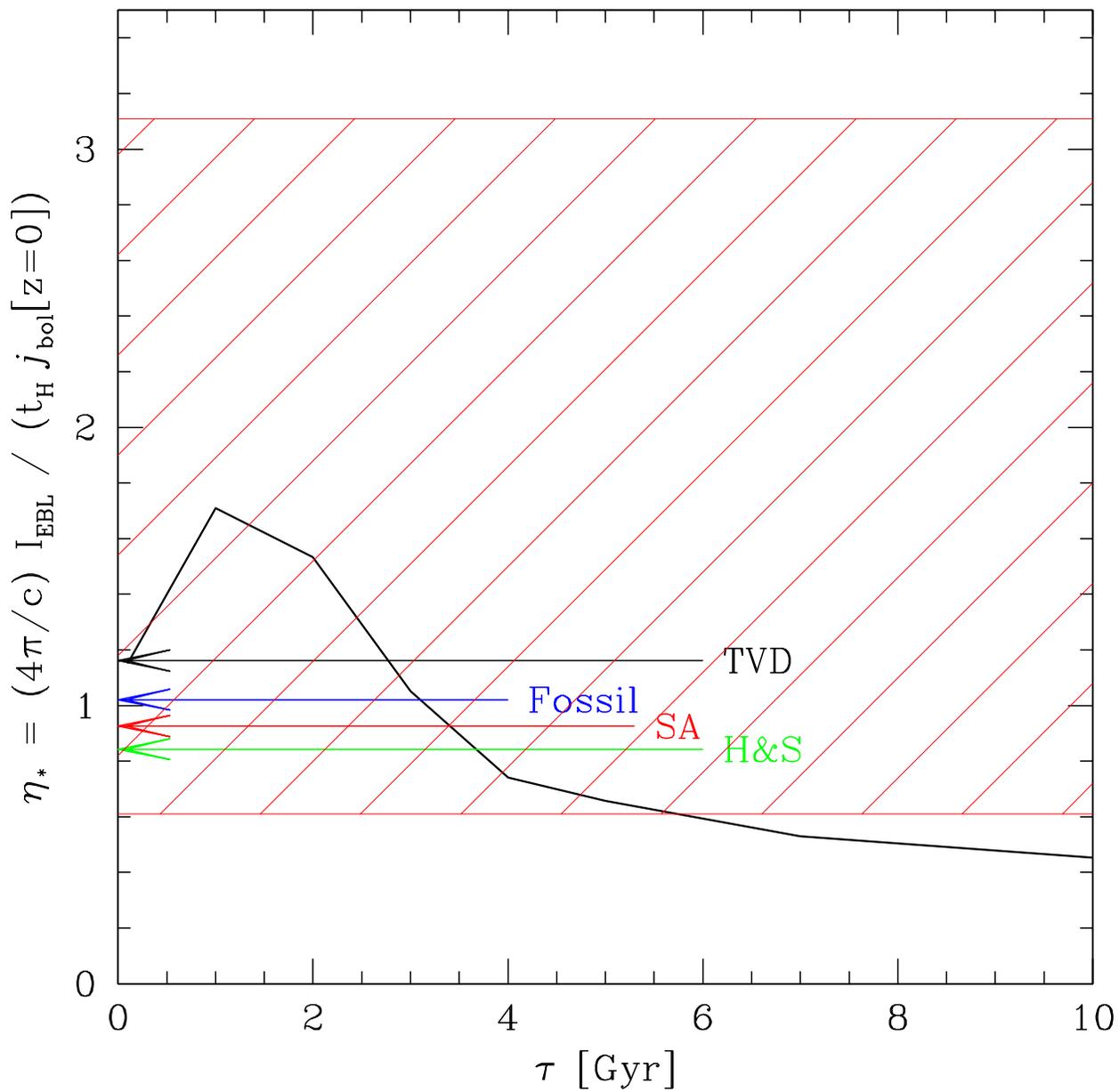}
\caption{
Parameter $\eta_*$ of Equation~(\ref{eq:eta}) computed for the
models used in this paper (shown by arrows).
The variation of $\eta_*$ is also shown
as a function of decay time-scale $\tau$ for the delayed 
exponential model of star formation. 
The shaded region shows the observationally allowed range.
}
\label{fig:eta}
\end{figure}

We may have another look at this problem by considering the ratio 
of the EBL to the bolometric luminosity density at $z=0$. 
We define the dimensionless parameter
\begin{equation}
\etastar \equiv \left(\frac{4\pi}{ct_H}\right) \frac{I_{EBL}}{\jbol}
= 7.293\times 10^6\  \frac{I_{EBL}\, 
[{\rm nW\, m^{-2}\, sr^{-1}}]}{\jbol\, [L_{\odot, {\rm bol}}\, \mpc^{-3}]}, 
\label{eq:eta}
\end{equation}
where $t_H \equiv 1/H_0$ is the Hubble time. 
This parameter compares the EBL with the luminosity 
density multiplied by the Hubble time, and
the advantage is that 
it is independent of uncertainties in the stellar IMF
provided that it does not vary as a function of time. 
From Figures~\ref{fig:bolum} and \ref{fig:ebl}, 
we compute $\etastar$, as shown in Figure~\ref{fig:eta}.  
The results are indicated by arrows, and 
the solid line shows $\etastar$ 
as a function of $\tau$ for the delayed exponential models. 
They are compared with the  
observation indicated by shades that 
ranges from $\etastar =0.61$ to 3.11 
from equations (\ref{eq:obsebl}) and (\ref{eq:jbol}). 
All models result in $\etastar = 1.0\pm0.2$, 
which is consistent with only the lower part of the 
apparently observed range. 
Therefore, the problem is IMF independent.


\section{Inconsistency among observations}
\label{sec:inconsistent}

The evidence we have discussed in the previous sections indicate
that we now have reasonable understanding of the cosmic star formation 
history.  There are, however, two items that do not fit well to 
the scenario.  They are the stellar mass density at $z\gtrsim 1$ and 
the EBL from direct measurements of sky.

We showed that the observationally derived stellar mass densities
at $z\ge 1$ are significantly smaller than the prediction of all the
models, while the models agree well with the observed SFR up to $z\sim 5$.
This means that the observational data of SFR and stellar mass density
are inconsistent with each other. 
The integration of SFR, model or observation, gives the correct 
stellar mass density at $z=0$, which is a strong constraint. 
Hence, this is clearly an observational problem that observations 
underestimated the stars locked in galaxies at $z\gtrsim 1$. 
Is it possible that the estimates of the stellar mass density are 
correct at all redshift and the SFR is mis-estimated? 
This is unlikely, because, to give the curve of the observed stellar 
mass density at $z\gtrsim 1$, SFR must decline sharply from $z\approx 0.5$ to 
higher redshift while the SFR density $z<0.5$ is substantially higher 
than is measured. 
Therefore, we may ascribe the problem to underestimates of the stellar 
mass density at $z\gtrsim 1$, presumably from poor understanding of the 
optical/NIR luminosity functions for sub-luminous galaxies..

The observation of EBL harbors another problem.  All model results  
are consistent with the EBL only when we take the value obtained by
integration of resolved sources, in {\it both} optical and infrared bands.
In the optical band, direct measurements \citep{Bernstein02a} 
give the EBL 3 times that 
from integration of resolved sources \citep{Madau00, Totani01a}. 
This factor becomes 5 in the infrared 
\citep[see][and references therein]{Hauser01}. 
The directly measured EBL in
either of the cases gives too bright a flux to be accounted for. 
This conclusion does not depend on cosmological theories, nor
on the IMF. It is a result of requiring consistencies between 
stellar mass density today and the light that has been emitted 
by those stars with subsidiary information concerning the 
star formation history inferred from the Madau diagram. 
Therefore this is also an observational problem possibly 
due to insufficient subtraction of the Zodiacal light
\citep[see also][]{Kneiske02, Mattila03}
\footnote{See Note Added in the end of this paper.}.
Finally there is an inconsistency between the observed value of
the luminosity density and the observed EBL for any plausible
star formation history. 

One may argue for the possibility that EBL in the infrared is
actually bright as a result of emission from high-redshift stars
which are not included in our modeling. 
We consider that this is unlikely, since we do not find at $z=0$ 
extra stars that were responsible for such emission at high-redshift, 
and the amount of stars at $z=0$ is well documented.  
If there were very massive stars such as Population III 
\citep[e.g.][]{Cambresy01, Matsumoto01, Cooray04, Kashlinsky04, 
Kashlinsky05a, Matsumoto05}, 
these objects would die quickly and will not be counted as stars 
at the present epoch.  However there are several difficulties 
with this hypothesis, such as much higher SFR at $z>7$ than at 
$z= 3-5$ \citep{Fernandez05}, possibility of over-enriching 
the intergalactic medium by metals \citep{Salvaterra03, Ricotti04a}, 
physically unrealistic absorption-corrected spectra of 
distant TeV blazars \citep{Dwek05a, Dwek05b}, and overproduction 
of soft X-ray background \citep{Madau05}.


\section{Conclusions}
\label{sec:discussion}

We have considered three physically motivated approaches:
one with {\it ab initio} theoretical models that are constrained 
basically by the observation of WMAP at $z\approx 1000$,
another with the Fossil model that uses theory of stellar    
evolution constrained by the SDSS observation at 
$z\approx 0.1$, and yet another purely from observations 
at $z=0-6$.  
We considered various physical quantities that witness
the star formation history: stellar mass density, luminosity density, 
bolometric luminosity density and EBL, and we found a general 
agreement among results from these three approaches 
within the errors the current observations indicate.
In particular, we found that the exceedingly simple Fossil model 
having two populations with parameters adjusted to the SDSS observation 
--- an update of the \citet{Searle73} model ---
provides a very effective description of the star
formation history predicted by {\it ab initio} theoretical models, 
and it also exhibits a good fit to the suite of tests we have applied. 

The only exception was in the stellar mass density at high redshift, 
where not only the Fossil model but all models predict higher 
values than the observationally derived ones.
The observational results do not look consistent 
with what are inferred from the empirical Madau plot.
This is not a matter of normalization of the star formation 
histories shown in Fig.~\ref{fig:sfr}, since the data 
on stellar mass density at $z=0$ are now well converged, 
and they are consistent with the integral of the SFR 
when the gas recycling is taken into account 
\citep[see also][]{Fukugita04}. 
If the dust extinction effect were significantly weaker 
than what we adopted here and the normalization 
of the SFR history were brought down in such a way that 
the integral of the SFR is consistent with the current 
data on stellar mass density at $z>1$, 
such a SFR history would be inconsistent with the 
measured stellar mass density at $z=0$.
Therefore, we regard the disagreement in the stellar mass density 
at $z>1$ between the models and observations 
as an observational problem. 

We also noted that the predictions of the EBL from the models 
all converged at the lower edge of the best current observations, 
giving $\EBL = 43\pm 7\,\eblunit$. 
This implies that the true EBL flux is presumably close to the value 
obtained by integrating over resolved sources. 

Before proceeding further, we should ask if an incorrect modeling
of the stellar IMF could be the culprit. The number and mass in 
stars less massive than $0.4\Msun$ are quite uncertain as they do 
not significantly affect observed spectral properties and their
contribution to dynamically determined mass estimates is obscured
by the uncertainties in the dark matter component.   
For our analysis we consistently adopted the Chabrier IMF 
($0.01-100\Msun$). 
Let us now suppose that, due to a higher than expected fraction of 
low-mass stars (as compared to the Chabrier IMF), the effective 
mass-to-light ratio values were all to be increased by a 
factor of $\mu$ ($\mu \approx 1.6$ for the Salpeter IMF; 
values of $\mu < 1$ 
are also permitted if there were a even stronger turn-over of 
the stellar number density with decreasing mass as compared to 
the Chabrier IMF).  This change would alter the theoretical models 
by reducing the EBL by a factor of about $\mu$.  The observed values 
of EBL would of course be unchanged. 
Thus, increasing the contribution of low-mass stars as compared to 
the Chabrier IMF would exacerbate the problem of under-predicting
the EBL for all the models. 
On the other hand, reducing the contribution of low-mass stars 
even more so than the Chabrier IMF would ease the disparity between 
the predicted and observed EBL.  But then the models would overpredict
the present-day luminosity density in various bands and disrupt 
the rough agreement with the observed values that we found in 
Table~\ref{table:params}.  
Therefore no simple adjustment of the stellar IMF can be made to 
remove all discrepancies. 
In addition, the parameter $\etastar$ was constructed 
(Equation~[\ref{eq:eta}]) in a fashion so as to avoid uncertainties
in the stellar IMF and it too shows a discrepancy between observed
and predicted value. 
Therefore it seems inevitable that the models of cosmic
star formation history that we considered in this paper cannot
account for a total optical-to-IR EBL of $\sim 100$ nW m$^{-2}$ sr$^{-1}$. 
We would suggest that the direct observation of the EBL in sky may
still be contaminated by local emissions.   
If future observations would converge to a higher EBL value, 
as the current direct EBL observations indicate, 
we would get into a trouble in modeling the history of galaxies: 
we may have to invoke additional components, such as mini black holes 
or population III stars with a top-heavy IMF that are not 
included in the current modeling 
as discussed in Section~\ref{sec:inconsistent}.

An interesting feature of our Fossil model is that
it predicts that the spheroid component formed
predominantly at $z\gtrsim 1.5$, 
and the disk component formed mostly at $z\ltsim 1$. 
The early formation of the spheroidal component 
is consistent with some recent direct 
numerical simulations \citep[e.g.,][]{Abadi03, Naab06b}. 
This agrees with the observational knowledge 
known for some time \citep[e.g.,][]{Fukugita96}
and the early work by \citet{Searle73} and predecessors. 
Although the current {\it ab initio} models cannot distinguish
between these two components, the Fossil model provides one 
example how the total population can be separated into
the two population, consistently with observations.
As a consequence of the early formation of the spheroidal
component, our models clearly indicate a relatively constant
star formation rate from $z=7$ to $z=2$.  It will be interesting
to see if this prediction is confirmed or refuted by 
ongoing observational programs. 

We also note that the rapid decline of  
the spheroid formation from $z=3$ to $z=1$ in the 
Fossil model works in favor of having a population of 
Extremely Red Objects \citep[EROs; e.g.][]{McCarthy04b} 
at $z=1$ as found by recent observations 
\citep[e.g.][]{Cimatti04, Glazebrook04, McCarthy04a}. 
The existence of 
these EROs were first regarded as a challenge to the 
hierarchical CDM models \citep{Som04}, but was later shown 
\citep{Nachos2, Nachos3} that the overall space density of EROs 
could be accounted for in cosmological hydro-simulations.
Having a high value of SFR density at high-redshift ($z\ge 4$) 
as our models (except the SA model) helps to resolve the 
issues with the existence of high-redshift EROs.

We conclude that the agreement among the three different approaches,  
as summarized in Table~\ref{table:params}, is encouraging, 
and that the Fossil model provides a simple effective description 
of cosmological star formation history with very few free parameters.  
Furthermore, the general agreement between the TVD, H\&S, and the
Fossil model suggests that the cosmological hydrodynamic simulation 
based on a cold dark matter model is providing a reasonably 
accurate picture of cosmological star formation history
without a fine tuning of input parameters.


\acknowledgments 

We thank Thorsten Naab, Alice Shapley, and Michael Strauss for 
useful comments on the manuscript. 
We are grateful to Alan Heavens and Armin Gabasch for 
providing us with the SFR data points in electronic form. 
This work was supported in part by grants 
NAG5-13381, AST 05-07521 and NNG05GK10G.
MF received support by Grant in Aid of the Ministry of Education
at the University of Tokyo and by the Monell Foundation at 
the Institute for Advanced Study.
The cosmological TVD hydrodynamic simulation was carried out
at the National Center for Supercomputing Applications (NCSA).

\vspace{1cm}

NOTE ADDED:\\
After the submission of our paper, \citet{Aharonian06} published 
an important upper limit to the NIR background from the
HESS observation of high energy gamma rays from two blazers,
H2356$-$309 and 1ES1101$-$232. The absence of absorption of
gamma rays excludes the high value of EBL inferred from NIR fluctuation
measurements, while the EBL from the source count is consistent
with the observation. This agrees with our conclusion derived
from the analysis given in Sect. 5.2. We also note that the source 
count recently measured with Spitzer \citep{Dole06}
agrees with the EBL we concluded.   


\appendix

\section{Mean metallicity of bulge}
\label{sec:bulge_metal}

We calculate the mass-weighted mean metallicity of 
the early type galaxies and the bulge component of 
spiral galaxies by
\begin{eqnarray}
\avg{Z} &=& \frac{\int Z(M)\, M\, \frac{dN}{dM}\, dM}{\int M\, \frac{dN}{\dd M}\, d M} \\
&=& \frac{\int Z(L)\, L\, \rp{\frac{M}{L}} \frac{dN}{dL}\, dL}{\int L\, \rp{\frac{M}{L}} \frac{dN}{dL}\, dL} \\
&=& \frac{\int Z(L)\, L^{1.14}\, \frac{dN}{d\log L}\, d\log L}{\int L^{1.14}\, \frac{dN}{d\log L}\, d\log L}, 
\label{eq:metal}
\end{eqnarray}
where $Z(M)$ and $Z(L)$ are the metallicity of the bulge component as 
a function of stellar mass $M$ and luminosity $L$. 
In equation~(\ref{eq:metal}) we used the scaling 
$(M/L) \propto L^{0.14}$ for the stellar mass-to-light ratio 
of late-type galaxies \citep{Bernardi03c, Vale04}.
We estimate the function $Z(L)$ from the relation 
between metallicity and velocity dispersion,
\begin{equation}
[Z/{\rm H}] = 0.53\, (\log \sigma - 2.173) + 0.15
\label{eq:funda}
\end{equation}
obtained by 
\citet{Nelan05}.
Using the Faber-Jackson relation \citep{Faber76},
\begin{equation}
\sigma = \frac{220}{\sqrt{2}}\rp{\frac{L}{L^*}}^{0.25}\ \kms, 
\label{eq:faber}
\end{equation}
we obtain
\begin{equation}
Z(L) = 0.13\, \log \rp{\frac{L}{L^*}} + 0.16.
\label{eq:ZL}
\end{equation}
We assume the same relation for bulges of late type galaxies.

We take $r$-band luminosity functions (LFs) of different morphological
types of galaxies from \citet{Nakamura03},
and scale down the luminosity of S0/a-Sb, Sbc-Sd galaxies 
by factors of of 0.40 and 0.24 (see Table~\ref{table:bulgetodisk}),  
respectively, to exclude the contribution from the disk component. 
The result of the integral (\ref{eq:metal}) is presented 
in Table~\ref{table:bulge}.

A caveat is that Equation~(\ref{eq:ZL}) may not be entirely 
appropriate for the bulges of late types galaxies, 
because the relation suggests
the metallicity of the bulge of the Milky Way would be 
super-solar (0.5 dex), while observations suggest sub-solar metallicity
(-0.5 dex) for bulge stars of the Milky Way \citep[e.g.,][]{Freeman02}.
We may have overestimated the mean metallicity for bulges of disk galaxies.


\section{Determination of the Normalization of the Fossil Model}
\label{sec:fossil_norm}

The disk and the bulge components of the Fossil model 
are normalized to the observed luminosity densities
$j_{K,tot} = 4.1\times 10^8\,h_{70}L_{\odot, K} \mpc^{-3}$ 
\citep{Bell03, Cole01} and 
$j_{B,tot} = 1.3\times 10^8\,h_{70}L_{\odot, B} \mpc^{-3}$ 
\citep{Fukugita04}. 
The luminosity densities are decomposed into
spheroids and disk components, as 
\begin{eqnarray}
j_{K,tot} &=& j_{K,b} + j_{K,d} = b_1\, \rhosb + d_1\, \rhosd, \\
\label{eq:jk}
j_{B,tot} &=& j_{B,b} + j_{B,d} = b_2\, \rhosb + d_2\, \rhosd
\label{eq:jb}
\end{eqnarray}
where the subscripts $b$ and $d$ stand for 
the spheroid and the disk, with the coefficients 
the inverse of the stellar mass-to-light ratios
$b_1 = (L_K/\Mstar)_b$, $b_2 = (L_B/\Mstar)_b$, 
$d_1 = (L_K/\Mstar)_d$, and $d_2 = (L_B/\Mstar)_d$. 
The BClib03 model give
$b_1 = (0.86)^{-1}=1.16$, $b_2=(5.59)^{-1}=0.18$, 
$d_1 = (0.55)^{-1}=1.82$, and $d_2 = (1.20)^{-1}=0.83$. 
We solve the equations and obtain
$\rhosb = 1.63\times 10^8\,\Msun\,\mpc^{-3}$ and 
$\rhosd = 1.21\times 10^8\,\Msun\,\mpc^{-3}$. 
The corresponding stellar mass density parameters are
$\Os = 0.00120$ (bulge) and $0.00089$ (disk) 
as presented in Table~\ref{table:fossil} and \ref{table:params}, 
which means the mean bulge-to-disk mass ratio $(M_b/M_d)=1.35$ 
(or equivalently $M_b:M_d=57\%: 43\%$). 
This bulge mass fraction is somewhat smaller than the one 
obtained by explicit summation over the galaxy sample,
but it is within the expected uncertainty of $\sim 20$\%. 
We note that this method does not work well if one uses
the $r$-band luminosity instead of $B$-band luminosity, because
$r$-band is not sensitive enough to young stars
that dominate the disk component. 



\newpage

\begin{deluxetable}{lcccccc}  
\tablecolumns{7}  
\tablewidth{0pc}  
\tablecaption{Bulge-disk luminosity decomposition based on SDSS data}
\tablehead{
\colhead{} & \colhead{E} & \colhead{S0} & \colhead{S0/a-Sb} & \colhead{Sbc-Sd}  & \colhead{Irr} & \colhead{$j_r$} 
}
\startdata
$j_r$\,\tablenotemark{a}............................ & 0.21 & 0.41 & 1.00 & 0.37 & 0.02 & 2.00 \cr
B/T\,\tablenotemark{b}.......................... & 1.00 & 0.62 & 0.40 & 0.24 & 0.00 & 0.95 \cr
D/T\,\tablenotemark{c}.......................... & 0.00 & 0.38 & 0.60 & 0.76 & 1.00 & 1.05 \cr
\enddata
\tablenotetext{a}{~$r$ band luminosity density contributed by each Hubble type 
of galaxies from \citet{Nakamura03}.}
\tablenotetext{b, c}{~Bulge-to-total and disk-to-total luminosity 
ratios for each type of galaxies from \citet{Ohama03}.}
\tablecomments{The last column gives 
the $r$ band luminosity densities for the bulge and disc components
in units of $10^8\,h L_{\odot, r} \mpc^{-3}$, which
yield $j_r(B)/j_r(D)=0.91$. 
Extinction corrections are applied to the disk component, using 
$A_R=\gamma_R \log (a/b)$, where $a/b$ is the axis ratio, 
and $\gamma_R = 1.15 + 1.88 (\log W_{20\%} - 2.5) = 1.24$  
at $\log W_{20\%} = 2.55$ for $R^* = -21.09$ from 
\citet[][see also \citet{Tully98}]{Sakai00}:
for $\log \avg{a/b}=0.22$ 
we obtain $\langle A_R\rangle=0.27$, 
where the difference between $R$ and $r$ is ignored. 
This leads us to the corrected bulge-to-disk ratio 
for the luminosity density,
$j_r(B)/j_r(D)|_{\rm corr} =j_r(B)/j_r(D)10^{0.4\avg{A_R}}  = 0.71$. 
We assign a 20\% error to this quantity. 
}
\label{table:bulgetodisk}
\end{deluxetable}

\begin{deluxetable}{cccccccccc}  
\tablecolumns{10}  
\tablewidth{0pc}  
\tablecaption{Properties of the Fossil model at $z=0$}
\tablehead{
\colhead{Population} & \colhead{$\Os$} & \colhead{$\tau$\,\tablenotemark{a}} & \colhead{$Z/Z_{\odot}$\tablenotemark{b}} & \colhead{$\avg{t_{\rm age}}$\,\tablenotemark{c}} & \colhead{$\ug$}  & \colhead{$\gr$} & \colhead{$(M_\star/L_B)$\,\tablenotemark{d}} & \colhead{$(M_\star/L_r)$\,\tablenotemark{e}} &  \colhead{$(M_\star/L_K)$\,\tablenotemark{f}}
}
\startdata
Bulge  & 0.00120 & 1.5 & 1.5 & 10.5 & 1.78 & 0.84 & 5.59 & 3.19 & 0.86 \cr
Disk   & 0.00089 & 4.5 & 0.8 &  7.0 & 0.94 & 0.43 & 1.20 & 1.18 & 0.55 \cr
\enddata
\tablenotetext{a}{~Characteristic time-scale in units of Gyr 
[defined in Equation~(\ref{eq:delayexp})].} 
\tablenotetext{b}{~Metallicity ($Z_{\odot}=0.02$).}
\tablenotetext{c}{~Mean age of stars in units of Gyr.}
\tablenotetext{d-f}{~Stellar-mass-to-light ratio in solar units for
the $B,r,K$-bands and the Chabrier IMF ($0.01 - 100\,\Msun$). 
Stellar masses include the remnants.}
\label{table:fossil}
\end{deluxetable}

\begin{deluxetable}{lcccccccccccc}  
\tabletypesize{\small}
\tablecolumns{13}  
\tablewidth{0pc}  
\tablecaption{Physical quantities at $z=0$ for the models used in the text}
\tablehead{
\colhead{Model} & \colhead{$\Os$\tablenotemark{\dagger}} & \colhead{$M_\star$\tablenotemark{\natural}} & \colhead{$M_{\rm rem}$\tablenotemark{\forall}}  & \colhead{$I_{EBL}$\tablenotemark{\ddag}} & \colhead{$\dot{\rhos}$\tablenotemark{\S}}  &  \colhead{$\jbol$\tablenotemark{a}} & \colhead{$j_U$\tablenotemark{b}} & \colhead{$j_B$\tablenotemark{c}}  & \colhead{$j_r$\tablenotemark{d}} & \colhead{$j_K$\tablenotemark{e}} & \colhead{$\ug$} & \colhead{$\gr$}  
}
\startdata
Fossil       & 0.0021 & 2.26 & 0.58 & 42 & 0.86  & 3.0 & 1.4 & 1.3 & 1.5 & 4.1 & 1.54 & 0.67 \cr
..... bulge & 0.00120 & 1.28 & 0.36 & 22 & 0.022 & 0.9 & 0.2 & 0.3 & 0.5 & 1.9 & 1.78 & 0.84 \cr
..... disk  & 0.00089 & 0.99 & 0.22 & 20 & 0.837 & 2.1 & 1.2 & 1.0 & 1.0 & 2.2 & 0.94 & 0.43 \cr
TVD          & 0.0028 & 3.06 & 0.80 & 51 & 1.02  & 3.2 & 1.5 & 1.4 & 1.8 & 5.1 & 1.20 & 0.61 \cr
H\&S         & 0.0021 & 2.23 & 0.56 & 37 & 1.30  & 3.2 & 1.7 & 1.5 & 1.6 & 4.1 & 1.00 & 0.50 \cr
SA           & 0.0023 & 2.56 & 0.61 & 47 & 1.37  & 3.7 & 2.0 & 1.7 & 2.0 & 4.9 & 1.04 & 0.52 \cr 
\hline
Consensus & 0.0023 & 2.52 & 0.65 & 43 & 1.06 & 3.1 & 1.5 & 1.4 & 1.6 & 4.4 & 1.25 & 0.59 \cr  
Model\tablenotemark{\P} & $\pm 0.0004$ & $\pm 0.47$ & $\pm 0.13$ & $\pm 7$ & $\pm 0.22$ & $\pm 0.2$ & $\pm 0.2$ & $\pm 0.1$ & $\pm 0.2$ & $\pm 0.6$ & $\pm 0.27$ & $\pm 0.09$ \cr
\hline
Observed & 0.0018 & --- & --- & $42 -$ & 0.5 & 3.2 & 1.4 & 1.2 & 1.27 & 3.4 & --- & --- \cr
range    & $-0.0039$ & --- & --- & 135   & $-1.6$ & $-5.0$ & $-1.8$ & $-2.0$ & $-1.33$ & $-5.5$ & --- & --- \cr  
\enddata
\tablenotetext{\dagger}{~Stellar mass density including the remnants for 
Chabrier IMF ($0.01 - 100\,\Msun$).}
\tablenotetext{\natural}{~Stellar mass density without the remnants 
(in $10^8\,\Msun\,\mpc^{-3}$).} 
\tablenotetext{\forall}{~Stellar remnant mass density 
(in $10^8\,\Msun\,\mpc^{-3}$).} 
\tablenotetext{\ddag}{~EBL intensity 
(in nW m$^{-2}$ sr$^{-1}$).} 
\tablenotetext{\S}{~SFR density 
(in $10^{-2}\,\Msun\,\yr^{-1}~{\rm Mpc}^{-3}$).} 
\tablenotetext{a}{~Bolometric luminosity density 
(in $10^8\,L_{\odot, {\rm bol}} \mpc ^{-3}$).} 
\tablenotetext{b-e}{~Luminosity density in 
$U,B,r,K$-bands (in $10^8\,L_{\odot, B} \mpc ^{-3}$). }
\tablenotetext{\P}{~Average of the Fossil, TVD, and H\&S models, 
with errors being the dispersion among the three models.}
\tablecomments{
Bolometric magnitudes $M_{B,\odot}=5.48$, $M_{K,\odot}=3.28$ 
and $M_{r, \odot}=4.76$ 
are adopted for the Sun.
Model luminosity densities are compared with observations:
$j_B = (1.7\pm 0.3)\times 10^8\,L_{\odot, B} \mpc^{-3}$ 
\citep[][]{Nag01b},
$j_B = (1.3\pm 0.1)\times 10^8\,L_{\odot, B} \mpc^{-3}$ 
\citep{Fukugita04}, 
$j_r = (1.3\pm 0.03)\times 10^8\,L_{\odot, r} \mpc^{-3}$ 
\citep[at $z=0.1$,][]{Blanton03a}, 
$j_K = 4.1^{+2.0}_{-0.6}\times 10^8\,L_{\odot, K} \mpc^{-3}$ 
\citep{Bell03}, $(5.0\pm 0.5)\times 10^8\,L_{\odot, K} \mpc^{-3}$
\citep{Kochanek01}, $j_{Ks} = (4.0\pm 0.6)\times 10^8\,L_{\odot, Ks} 
\mpc^{-3}$ \citep{Cole01}. 
All values assume $h=0.7$. 
}
\label{table:params}
\end{deluxetable}

\begin{deluxetable}{lcccccccc}  
\tablecolumns{3}  
\tablewidth{0pc}  
\tablecaption{Mean metallicity of spheroids}
\tablehead{
\colhead{Type} & \colhead{$\avg{Z/\Zsun}$ \tablenotemark{a}} & 
\colhead{$j_r\,[10^8\,h\,L_{r,\odot} \mpc^{-3}$] \tablenotemark{b}} 
}
\startdata
E-S0 .........................................  & 1.44 & 0.70  \cr
S0/a--Sb ................................... & 1.15 & 0.39  \cr
Sbc--Sd ..................................... & 1.15 & 0.082 \cr
\hline
Total ......................................... & 1.33 & 1.17  \cr 
\enddata
\tablenotetext{a}{~Mass weighted mean metallicity.}
\tablenotetext{b}{~$r$-band luminosity density from the spheroid component.}
\label{table:bulge}
\end{deluxetable}


\newpage 

\begin{thebibliography}{118}
\expandafter\ifx\csname natexlab\endcsname\relax\def\natexlab#1{#1}\fi

\bibitem[{{Abadi} {et~al.}(2003){Abadi}, {Navarro}, {Steinmetz}, \&
  {Eke}}]{Abadi03}
{Abadi}, M.~G., {Navarro}, J.~F., {Steinmetz}, M., \& {Eke}, V.~R. 2003, \apj,
  591, 499

\bibitem[{{Aharonian} {et~al.}(2006){Aharonian}, {Akhperjanian}, {Bazer-Bachi},
  {Beilicke}, {Benbow}, {Berge}, {Bernl{\"o}hr}, {Boisson},
  {et~al.}}]{Aharonian06}
{Aharonian}, F., {Akhperjanian}, A.~G., {Bazer-Bachi}, A.~R., {Beilicke}, M.,
  {Benbow}, W., {Berge}, D., {Bernl{\"o}hr}, K., {Boisson}, C., {et~al.} 2006,
  \nat, 440, 1018

\bibitem[{{Arnett}(1996)}]{Arnett}
{Arnett}, D. 1996, {Supernovae and nucleosynthesis. an investigation of the
  history of matter, from the Big Bang to the present} (Princeton series in
  astrophysics, Princeton, NJ: Princeton University Press, 1996)

\bibitem[{Ascasibar {et~al.}(2002)Ascasibar, Yepes, Gottl\"{o}ber, \&
  M\"{u}ller}]{Ascasibar02}
Ascasibar, Y., Yepes, G., Gottl\"{o}ber, S., \& M\"{u}ller, V. 2002, A\&A, 387,
  396

\bibitem[{Baade(1944)}]{Baade44}
Baade, W. 1944, ApJ, 100, 137

\bibitem[{Baldry {et~al.}(2004)Baldry, Glazebrook, Brinkmann, Ivezic, Lupton,
  Nichol, \& Szalay}]{Baldry04}
Baldry, I.~K., Glazebrook, K., Brinkmann, J., Ivezic, Z., Lupton, R.~H.,
  Nichol, R.~C., \& Szalay, A.~S. 2004, ApJ, 600, 681

\bibitem[{Balogh {et~al.}(2004)Balogh, Baldry, Nichol, Miller, Bower, \&
  Glazebrook}]{Balogh04}
Balogh, M.~L., Baldry, I.~K., Nichol, R., Miller, C., Bower, R., \& Glazebrook,
  K. 2004, ApJ, 615, L101

\bibitem[{Barger {et~al.}(2000)Barger, Cowie, \& Richards}]{Barger00}
Barger, A.~J., Cowie, L.~L., \& Richards, E.~A. 2000, AJ, 119, 2092

\bibitem[{Bell {et~al.}(2003)Bell, McIntosh, Katz, \& Weinberg}]{Bell03}
Bell, E.~F., McIntosh, D.~H., Katz, N., \& Weinberg, M.~D. 2003, ApJ, 149, 289

\bibitem[{{Bernardi} {et~al.}(2003){Bernardi}, {Sheth}, {Annis}, {Burles},
  {Eisenstein}, {Finkbeiner}, {Hogg}, {Lupton}, {et~al.}}]{Bernardi03c}
{Bernardi}, M., {Sheth}, R.~K., {Annis}, J., {Burles}, S., {Eisenstein}, D.~J.,
  {Finkbeiner}, D.~P., {Hogg}, D.~W., {Lupton}, R.~H., {et~al.} 2003, \aj, 125,
  1866

\bibitem[{{Bernstein} {et~al.}(2002){Bernstein}, {Freedman}, \&
  {Madore}}]{Bernstein02a}
{Bernstein}, R.~A., {Freedman}, W.~L., \& {Madore}, B.~F. 2002, \apj, 571, 56

\bibitem[{Blanton {et~al.}(2003{\natexlab{a}})Blanton, Hogg, Bahcall,
  {et~al.}}]{Blanton03a}
Blanton, M.~R., Hogg, D.~W., Bahcall, N.~A., {et~al.} 2003{\natexlab{a}}, ApJ,
  592, 819

\bibitem[{Blanton {et~al.}(2003{\natexlab{b}})}]{Blanton03b}
Blanton, M.~R. {et~al.} 2003{\natexlab{b}}, ApJ, 594, 186

\bibitem[{{Bond} {et~al.}(1991){Bond}, {Cole}, {Efstathiou}, \&
  {Kaiser}}]{Bond91}
{Bond}, J.~R., {Cole}, S., {Efstathiou}, G., \& {Kaiser}, N. 1991, \apj, 379,
  440

\bibitem[{{Bouch{\'e}} {et~al.}(2006){Bouch{\'e}}, {Lehnert}, \&
  {P{\'e}roux}}]{Bouche06}
{Bouch{\'e}}, N., {Lehnert}, M.~D., \& {P{\'e}roux}, C. 2006, \mnras, 367, L16

\bibitem[{Bouwens {et~al.}(2005)Bouwens, Illingworth, Blakeslee, \&
  Franx}]{Bouwens05}
Bouwens, R.~J., Illingworth, G.~D., Blakeslee, J.~P., \& Franx, M. 2005, ApJ,
  accepted (astro-ph/0509641)

\bibitem[{Bouwens {et~al.}(2004)Bouwens, Illingworth, Thompson, Blakeslee,
  Dickinson, Broadhurst, Eisenstein, Fan, {et~al.}}]{Bouwens04a}
Bouwens, R.~J., Illingworth, G.~D., Thompson, R.~I., Blakeslee, J.~P.,
  Dickinson, M.~E., Broadhurst, T.~J., Eisenstein, D.~J., Fan, X., {et~al.}
  2004, ApJ, 606, L25

\bibitem[{Brinchmann {et~al.}(2004)Brinchmann, Charlot, White, Tremonti,
  Kauffmann, Heckman, \& Brinkmann}]{Brinchmann04}
Brinchmann, J., Charlot, S., White, S. D.~M., Tremonti, C., Kauffmann, G.,
  Heckman, T., \& Brinkmann, J. 2004, MNRAS, 351, 1151

\bibitem[{Brinchmann \& Ellis(2000)}]{Brinchmann00}
Brinchmann, J. \& Ellis, R. 2000, ApJ, 536, L77

\bibitem[{Bruzual \& Charlot(2003)}]{BClib03}
Bruzual, G. \& Charlot, S. 2003, MNRAS, 344, 1000

\bibitem[{Bunker {et~al.}(2004)Bunker, Stanway, Ellis, \& McMahon}]{Bunker04}
Bunker, A.~J., Stanway, E.~R., Ellis, R.~S., \& McMahon, R.~G. 2004, MNRAS,
  355, 374

\bibitem[{Cambresy {et~al.}(2001)Cambresy, Reach, Beichman, \&
  Jarrett}]{Cambresy01}
Cambresy, L., Reach, W.~T., Beichman, C.~A., \& Jarrett, T.~H. 2001, ApJ, 555,
  563

\bibitem[{{Cen} {et~al.}(2005){Cen}, {Nagamine}, \& {Ostriker}}]{Cen05}
{Cen}, R., {Nagamine}, K., \& {Ostriker}, J.~P. 2005, \apj, 635, 86

\bibitem[{Cen \& Ostriker(1993)}]{CO93}
Cen, R. \& Ostriker, J.~P. 1993, ApJ, 417, 404

\bibitem[{Chabrier(2003)}]{Chab03a}
Chabrier, G. 2003, ApJ, 586, L133

\bibitem[{Cimatti {et~al.}(2004)Cimatti, Daddi, Renzini, Cassata, Vanzella,
  Pozzetti, Cristiani, Fontana, {et~al.}}]{Cimatti04}
Cimatti, A., Daddi, E., Renzini, A., Cassata, P., Vanzella, E., Pozzetti, L.,
  Cristiani, S., Fontana, A., {et~al.} 2004, Nature, 430, 184

\bibitem[{Cole {et~al.}(2000)Cole, Lacey, Baugh, \& Frenk}]{Cole00}
Cole, S., Lacey, C.~G., Baugh, C.~M., \& Frenk, C.~S. 2000, MNRAS, 319, 168

\bibitem[{Cole {et~al.}(2001)Cole, Norberg, Baugh, Frenk, Bland-Hawthorn,
  Bridges, Cannon, Colless, Collins, {et~al.}}]{Cole01}
Cole, S., Norberg, P., Baugh, C.~M., Frenk, C.~S., Bland-Hawthorn, J., Bridges,
  T., Cannon, R., Colless, M., Collins, C., {et~al.} 2001, MNRAS, 326, 255

\bibitem[{{Cooray} {et~al.}(2004){Cooray}, {Bock}, {Keatin}, {Lange}, \&
  {Matsumoto}}]{Cooray04}
{Cooray}, A., {Bock}, J.~J., {Keatin}, B., {Lange}, A.~E., \& {Matsumoto}, T.
  2004, \apj, 606, 611

\bibitem[{Cowie {et~al.}(1999)Cowie, Songaila, \& Barger}]{Cowie99}
Cowie, L.~L., Songaila, A., \& Barger, A.~J. 1999, AJ, 118, 603

\bibitem[{Dickinson {et~al.}(2003)Dickinson, Stern, Giavalisco, Ferguson,
  Tsvetanov, Chornock, Cristiani, Dawson, {et~al.}}]{Dickinson03b}
Dickinson, M., Stern, D., Giavalisco, M., Ferguson, H.~C., Tsvetanov, Z.,
  Chornock, R., Cristiani, S., Dawson, S., {et~al.} 2003, ApJ, 600, L99

\bibitem[{{Dole} {et~al.}(2006){Dole}, {Lagache}, {Puget}, {Caputi},
  {Fern{\'a}ndez-Conde}, {Le Floc'h}, {Papovich}, {P{\'e}rez-Gonz{\'a}lez},
  {et~al.}}]{Dole06}
{Dole}, H., {Lagache}, G., {Puget}, J.-L., {Caputi}, K.~I.,
  {Fern{\'a}ndez-Conde}, N., {Le Floc'h}, E., {Papovich}, C.,
  {P{\'e}rez-Gonz{\'a}lez}, P.~G., {et~al.} 2006, \aap, 451, 417

\bibitem[{{Dunne} {et~al.}(2003){Dunne}, {Eales}, \& {Edmunds}}]{Dunne03}
{Dunne}, L., {Eales}, S.~A., \& {Edmunds}, M.~G. 2003, \mnras, 341, 589

\bibitem[{{Dwek} {et~al.}(2005{\natexlab{a}}){Dwek}, {Arendt}, \&
  {Krennrich}}]{Dwek05b}
{Dwek}, E., {Arendt}, R.~G., \& {Krennrich}, F. 2005{\natexlab{a}}, \apj, 635,
  784

\bibitem[{{Dwek} {et~al.}(2005{\natexlab{b}}){Dwek}, {Krennrich}, \&
  {Arendt}}]{Dwek05a}
{Dwek}, E., {Krennrich}, F., \& {Arendt}, R.~G. 2005{\natexlab{b}}, \apj, 634,
  155

\bibitem[{{Faber} \& {Jackson}(1976)}]{Faber76}
{Faber}, S.~M. \& {Jackson}, R.~E. 1976, \apj, 204, 668

\bibitem[{{Fernandez} \& {Komatsu}(2005)}]{Fernandez05}
{Fernandez}, E. \& {Komatsu}, E. 2005, astro-ph/0508174

\bibitem[{Fontana {et~al.}(2004)Fontana, Pozzetti, Donnarumma, Renzini,
  Cimatti, Zamorani, Menci, Daddi, {et~al.}}]{Fontana04}
Fontana, A., Pozzetti, L., Donnarumma, I., Renzini, A., Cimatti, A., Zamorani,
  G., Menci, N., Daddi, E., {et~al.} 2004, A\&A, 424, 23

\bibitem[{{Freeman} \& {Bland-Hawthorn}(2002)}]{Freeman02}
{Freeman}, K. \& {Bland-Hawthorn}, J. 2002, \araa, 40, 487

\bibitem[{{Fukugita} {et~al.}(1996){Fukugita}, {Hogan}, \&
  {Peebles}}]{Fukugita96}
{Fukugita}, M., {Hogan}, C.~J., \& {Peebles}, P.~J.~E. 1996, \nat, 381, 489

\bibitem[{Fukugita \& Peebles(2004)}]{Fukugita04}
Fukugita, M. \& Peebles, P. J.~E. 2004, ApJ, 616, 643

\bibitem[{Gabasch {et~al.}(2004)Gabasch, Salvato, Saglia, Bender,
  {et~al.}}]{Gabasch04}
Gabasch, A., Salvato, M., Saglia, R.~P., Bender, R., {et~al.} 2004, ApJ, 616,
  L83

\bibitem[{Gallego {et~al.}(1995)Gallego, Zamorano, Aragon-Salamanca, \&
  Rego}]{Gallego95}
Gallego, J., Zamorano, J., Aragon-Salamanca, A., \& Rego, M. 1995, ApJ, 459,
  L43

\bibitem[{Giavalisco {et~al.}(2004)Giavalisco, Dickinson, Ferguson,
  Ravindranath, Kretchmer, Moustakas, Madau, Fall, {et~al.}}]{Giavalisco04}
Giavalisco, M., Dickinson, M., Ferguson, H.~C., Ravindranath, S., Kretchmer,
  C., Moustakas, L.~A., Madau, P., Fall, M., {et~al.} 2004, ApJ, 600, L103

\bibitem[{Glazebrook {et~al.}(2004)Glazebrook, Abraham, McCarthy, Savaglio,
  Chen, Crampton, Murowinski, Jorgensen, {et~al.}}]{Glazebrook04}
Glazebrook, K., Abraham, R., McCarthy, P., Savaglio, S., Chen, H.-W., Crampton,
  D., Murowinski, R., Jorgensen, I., {et~al.} 2004, Nature, 430, 181

\bibitem[{Hauser \& Dwek(2001)}]{Hauser01}
Hauser, M.~G. \& Dwek, E. 2001, ARA\&A, 39, 249

\bibitem[{Heavens {et~al.}(2000)Heavens, Jimenez, \& Lahav}]{Heavens00}
Heavens, A.~F., Jimenez, R., \& Lahav, O. 2000, MNRAS, 317, 965

\bibitem[{Heavens {et~al.}(2004)Heavens, Panter, Jimenez, \&
  Dunlop}]{Heavens04}
Heavens, A.~F., Panter, B., Jimenez, R., \& Dunlop, J. 2004, Nature, 428, 625

\bibitem[{Hernquist \& Springel(2003)}]{Her03}
Hernquist, L. \& Springel, V. 2003, MNRAS, 341, 1253

\bibitem[{{Hopkins}(2004)}]{Hopkins04}
{Hopkins}, A.~M. 2004, \apj, 615, 209

\bibitem[{Hopkins {et~al.}(2000)Hopkins, Connolly, \& Szalay}]{Hopkins00}
Hopkins, A.~M., Connolly, A.~J., \& Szalay, A.~S. 2000, AJ, 120, 2843

\bibitem[{Iwata {et~al.}(2003)Iwata, Ohta, Tamura, Ando, Wada, Watanabe,
  Akiyama, \& Aoki}]{Iwata03}
Iwata, I., Ohta, K., Tamura, N., Ando, M., Wada, S., Watanabe, C., Akiyama, M.,
  \& Aoki, K. 2003, PASJ, 55, 415

\bibitem[{{Kashlinsky}(2005)}]{Kashlinsky05a}
{Kashlinsky}, A. 2005, \physrep, 409, 361

\bibitem[{Kashlinsky {et~al.}(2004)Kashlinsky, Arendt, Gardner, Mather, \&
  Moseley}]{Kashlinsky04}
Kashlinsky, A., Arendt, R., Gardner, J.~P., Mather, J.~C., \& Moseley, S.~H.
  2004, ApJ, 608, 1

\bibitem[{{Kauffmann} {et~al.}(1993){Kauffmann}, {White}, \&
  {Guiderdoni}}]{Kau93}
{Kauffmann}, G., {White}, S.~D.~M., \& {Guiderdoni}, B. 1993, \mnras, 264, 201

\bibitem[{Kauffmann {et~al.}(2003)}]{Kau03b}
Kauffmann, G. {et~al.} 2003, MNRAS, 341, 54

\bibitem[{{Kneiske} {et~al.}(2002){Kneiske}, {Mannheim}, \&
  {Hartmann}}]{Kneiske02}
{Kneiske}, T.~M., {Mannheim}, K., \& {Hartmann}, D.~H. 2002, \aap, 386, 1

\bibitem[{Kochanek {et~al.}(2001)Kochanek, Pahre, Falco, Huchra,
  {et~al.}}]{Kochanek01}
Kochanek, C.~S., Pahre, M.~A., Falco, E.~E., Huchra, J.~P., {et~al.} 2001, ApJ,
  560, 566

\bibitem[{{Lacey} \& {Cole}(1993)}]{Lacey93}
{Lacey}, C. \& {Cole}, S. 1993, \mnras, 262, 627

\bibitem[{{Lanzetta} {et~al.}(2002){Lanzetta}, {Yahata}, {Pascarelle}, {Chen},
  \& {Fern{\'a}ndez-Soto}}]{Lanzetta02}
{Lanzetta}, K.~M., {Yahata}, N., {Pascarelle}, S., {Chen}, H.-W., \&
  {Fern{\'a}ndez-Soto}, A. 2002, \apj, 570, 492

\bibitem[{Lilly {et~al.}(1996)Lilly, F\`{e}vre, Hammer, \& Crampton}]{Lilly96}
Lilly, S.~J., F\`{e}vre, O.~L., Hammer, F., \& Crampton, D. 1996, ApJ, 460, L1

\bibitem[{Madau {et~al.}(1996)Madau, Ferguson, Dickinson, Giavalisco, Steidel,
  \& Fruchter}]{Madau96}
Madau, P., Ferguson, H.~C., Dickinson, E.~D., Giavalisco, M., Steidel, C.~C.,
  \& Fruchter, A. 1996, MNRAS, 283, 1388

\bibitem[{Madau \& Pozzetti(2000)}]{Madau00}
Madau, P. \& Pozzetti, L. 2000, MNRAS, 312, L9

\bibitem[{Madau {et~al.}(1998)Madau, Pozzetti, \& Dickinson}]{Madau98}
Madau, P., Pozzetti, L., \& Dickinson, M. 1998, ApJ, 498, 106

\bibitem[{Madau \& Silk(2005)}]{Madau05}
Madau, P. \& Silk, J. 2005, MNRAS, 359, L37

\bibitem[{Matsumoto(2001)}]{Matsumoto01}
Matsumoto, M. 2001, in "The extragalactic infrared background and its
  cosmological implications", IAU Symposium 204, ed. M.~Harwit \& M.~G. Hauser,
  101

\bibitem[{Matsumoto {et~al.}(2005)Matsumoto, Matsuura, Murakami, Tanaka,
  {et~al.}}]{Matsumoto05}
Matsumoto, T., Matsuura, S., Murakami, H., Tanaka, M., {et~al.} 2005, ApJ, 626,
  31

\bibitem[{{Mattila}(2003)}]{Mattila03}
{Mattila}, K. 2003, \apj, 591, 119

\bibitem[{McCarthy(2004)}]{McCarthy04b}
McCarthy, P.~J. 2004, ARA\&A, 42, 477

\bibitem[{McCarthy {et~al.}(2004)McCarthy, Le~Borgne, Crampton, Chen, Abraham,
  Glazebrook, Savaglio, Carlberg, {et~al.}}]{McCarthy04a}
McCarthy, P.~J., Le~Borgne, D., Crampton, D., Chen, H.-W., Abraham, R.~G.,
  Glazebrook, K., Savaglio, S., Carlberg, R.~G., {et~al.} 2004, ApJ, 614, L9

\bibitem[{{Naab} {et~al.}(2005){Naab}, {Johansson}, {Efstathiou}, \&
  {Ostriker}}]{Naab06b}
{Naab}, T., {Johansson}, P.~H., {Efstathiou}, G., \& {Ostriker}, J.~P. 2005,
  astro-ph/0512235

\bibitem[{{Naab} \& {Ostriker}(2006)}]{Naab06a}
{Naab}, T. \& {Ostriker}, J.~P. 2006, \mnras, 366, 899

\bibitem[{{Nagamine} {et~al.}(2006){Nagamine}, {Cen}, {Furlanetto},
  {Hernquist}, {Night}, {Ostriker}, \& {Ouchi}}]{Nag06a}
{Nagamine}, K., {Cen}, R., {Furlanetto}, S.~R., {Hernquist}, L., {Night}, C.,
  {Ostriker}, J.~P., \& {Ouchi}, M. 2006, New Astronomy Review, 50, 29

\bibitem[{Nagamine {et~al.}(2004)Nagamine, Cen, Hernquist, Ostriker, \&
  Springel}]{Nachos1}
Nagamine, K., Cen, R., Hernquist, L., Ostriker, J.~P., \& Springel, V. 2004,
  ApJ, 610, 45

\bibitem[{Nagamine {et~al.}(2005{\natexlab{a}})Nagamine, Cen, Hernquist,
  Ostriker, \& Springel}]{Nachos3}
---. 2005{\natexlab{a}}, ApJ, 627, 608

\bibitem[{Nagamine {et~al.}(2005{\natexlab{b}})Nagamine, Cen, Hernquist,
  Ostriker, \& Springel}]{Nachos2}
---. 2005{\natexlab{b}}, ApJ, 618, 23

\bibitem[{Nagamine {et~al.}(2000)Nagamine, Cen, \& Ostriker}]{Nag00}
Nagamine, K., Cen, R., \& Ostriker, J.~P. 2000, ApJ, 541, 25

\bibitem[{Nagamine {et~al.}(2001{\natexlab{a}})Nagamine, Fukugita, Cen, \&
  Ostriker}]{Nag01b}
Nagamine, K., Fukugita, M., Cen, R., \& Ostriker, J.~P. 2001{\natexlab{a}},
  MNRAS, 327, L10

\bibitem[{Nagamine {et~al.}(2001{\natexlab{b}})Nagamine, Fukugita, Cen, \&
  Ostriker}]{Nag01a}
---. 2001{\natexlab{b}}, ApJ, 558, 497

\bibitem[{{Nakamura} {et~al.}(2004){Nakamura}, {Fukugita}, {Brinkmann}, \&
  {Schneider}}]{Nakamura04}
{Nakamura}, O., {Fukugita}, M., {Brinkmann}, J., \& {Schneider}, D.~P. 2004,
  \aj, 127, 2511

\bibitem[{{Nakamura} {et~al.}(2003){Nakamura}, {Fukugita}, {Yasuda}, {Loveday},
  {Brinkmann}, {Schneider}, {Shimasaku}, \& {SubbaRao}}]{Nakamura03}
{Nakamura}, O., {Fukugita}, M., {Yasuda}, N., {Loveday}, J., {Brinkmann}, J.,
  {Schneider}, D.~P., {Shimasaku}, K., \& {SubbaRao}, M. 2003, \aj, 125, 1682

\bibitem[{{Nelan} {et~al.}(2005){Nelan}, {Smith}, {Hudson}, {Wegner}, {Lucey},
  {Moore}, {Quinney}, \& {Suntzeff}}]{Nelan05}
{Nelan}, J.~E., {Smith}, R.~J., {Hudson}, M.~J., {Wegner}, G.~A., {Lucey},
  J.~R., {Moore}, S.~A.~W., {Quinney}, S.~J., \& {Suntzeff}, N.~B. 2005, \apj,
  632, 137

\bibitem[{Norman {et~al.}(2004)Norman, Ptak, Hornschemeier, Hasinger, Bergeron,
  Comastri, Giacconi, Gilli, {et~al.}}]{Norman04}
Norman, C., Ptak, A., Hornschemeier, A., Hasinger, G., Bergeron, J., Comastri,
  A., Giacconi, R., Gilli, R., {et~al.} 2004, ApJ, 607, 721

\bibitem[{Ohama(2003)}]{Ohama03}
Ohama, N. 2003, PhD thesis, University of Tokyo

\bibitem[{Oort(1926)}]{Oort26}
Oort, J.~H. 1926, PGro, 40, 1

\bibitem[{Ostriker \& Steinhardt(1995)}]{Ostriker95}
Ostriker, J.~P. \& Steinhardt, P.~J. 1995, Nature, 377, 600

\bibitem[{Ouchi {et~al.}(2004a)Ouchi, Shimasaku, Furusawa, Miyazaki, Doi,
  Hamabe, Hayashino, Kimura, {et~al.}}]{Ouchi04a}
Ouchi, M., Shimasaku, K., Furusawa, H., Miyazaki, M., Doi, M., Hamabe, M.,
  Hayashino, T., Kimura, M., {et~al.} 2004a, ApJ, 611, 660

\bibitem[{Panter {et~al.}(2004)Panter, Heavens, \& Jimenez}]{Panter04}
Panter, B., Heavens, A.~F., \& Jimenez, R. 2004, MNRAS, 355, 764

\bibitem[{Pascual {et~al.}(2001)Pascual, Gallego, Arag\'{o}n-Salamanca, \&
  Zamorano}]{Pascual01}
Pascual, S., Gallego, J., Arag\'{o}n-Salamanca, A., \& Zamorano, J. 2001, A\&A,
  379, 798

\bibitem[{Perlmutter {et~al.}(1998)}]{Perlmutter98}
Perlmutter, S. {et~al.} 1998, Nature, 391, 51

\bibitem[{{Prochaska} {et~al.}(2006){Prochaska}, {O'Meara}, {Herbert-Fort},
  {Burles}, {Prochter}, \& {Bernstein}}]{Pro06}
{Prochaska}, J.~X., {O'Meara}, J.~M., {Herbert-Fort}, S., {Burles}, S.,
  {Prochter}, G.~E., \& {Bernstein}, R.~A. 2006, submitted, astro-ph/0606573

\bibitem[{{Reddy} {et~al.}(2005){Reddy}, {Erb}, {Steidel}, {Shapley},
  {Adelberger}, \& {Pettini}}]{Reddy05}
{Reddy}, N.~A., {Erb}, D.~K., {Steidel}, C.~C., {Shapley}, A.~E., {Adelberger},
  K.~L., \& {Pettini}, M. 2005, \apj, 633, 748

\bibitem[{{Reddy} \& {Steidel}(2004)}]{Reddy04}
{Reddy}, N.~A. \& {Steidel}, C.~C. 2004, \apjl, 603, L13

\bibitem[{{Ricotti} \& {Ostriker}(2004)}]{Ricotti04a}
{Ricotti}, M. \& {Ostriker}, J.~P. 2004, \mnras, 350, 539

\bibitem[{Riess {et~al.}(1998)}]{Riess98}
Riess, A.~G. {et~al.} 1998, AJ, 116, 1009

\bibitem[{Robin {et~al.}(2003)Robin, Reyl\'{e}, Derri\'{e}re, \&
  Picaud}]{Robin03}
Robin, A.~C., Reyl\'{e}, C., Derri\'{e}re, S., \& Picaud, S. 2003, A\&A, 409,
  523

\bibitem[{Rocha-Pinto {et~al.}(2000)Rocha-Pinto, Scalo, Maciel, \&
  Flynn}]{Rocha00}
Rocha-Pinto, H.~J., Scalo, J., Maciel, W.~J., \& Flynn, C. 2000, A\&A, 358, 869

\bibitem[{Rudnick {et~al.}(2003)Rudnick, Rix, Franx, Labbe, Blanton, Daddi,
  F\"{o}rster, Natascha, {et~al.}}]{Rudnick03}
Rudnick, G., Rix, H.-W., Franx, M., Labbe, I., Blanton, M., Daddi, E.,
  F\"{o}rster, S., Natascha, M., {et~al.} 2003, ApJ, 599, 847

\bibitem[{Sakai {et~al.}(2000)Sakai, Mould, Hughes, Huchra, {et~al.}}]{Sakai00}
Sakai, S., Mould, J.~R., Hughes, S. M.~G., Huchra, J.~P., {et~al.} 2000, ApJ,
  529, 698

\bibitem[{{Salvaterra} \& {Ferrara}(2003)}]{Salvaterra03}
{Salvaterra}, R. \& {Ferrara}, A. 2003, \mnras, 339, 973

\bibitem[{{Searle} {et~al.}(1973){Searle}, {Sargent}, \& {Bagnuolo}}]{Searle73}
{Searle}, L., {Sargent}, W.~L.~W., \& {Bagnuolo}, W.~G. 1973, \apj, 179, 427

\bibitem[{Somerville {et~al.}(2004)Somerville, Moustakas, Mobasher, Gardner,
  Cimatti, Conselice, Daddi, Dahlen, {et~al.}}]{Som04}
Somerville, R.~S., Moustakas, L.~A., Mobasher, B., Gardner, J.~P., Cimatti, A.,
  Conselice, C., Daddi, E., Dahlen, T., {et~al.} 2004, ApJ, 600, L135

\bibitem[{Somerville {et~al.}(2001)Somerville, Primack, \& Faber}]{Som01}
Somerville, R.~S., Primack, J.~R., \& Faber, S.~M. 2001, MNRAS, 320, 504

\bibitem[{Spergel {et~al.}(2003)Spergel, Verde, Peiris, Komatsu, Nolta,
  Bennett, Halpern, Hinshaw, {et~al.}}]{Spergel03}
Spergel, D., Verde, L., Peiris, H.~V., Komatsu, E., Nolta, M.~R., Bennett,
  C.~L., Halpern, M., Hinshaw, G., {et~al.} 2003, ApJS, 148, 175

\bibitem[{Springel \& Hernquist(2003{\natexlab{a}})}]{SH03a}
Springel, V. \& Hernquist, L. 2003{\natexlab{a}}, MNRAS, 339, 289

\bibitem[{Springel \& Hernquist(2003{\natexlab{b}})}]{SH03b}
---. 2003{\natexlab{b}}, MNRAS, 339, 312

\bibitem[{Steidel {et~al.}(1999)Steidel, Adelberger, Giavalisco, Dickinson, \&
  Pettini}]{Steidel99}
Steidel, C.~C., Adelberger, K.~L., Giavalisco, M., Dickinson, M., \& Pettini,
  M. 1999, ApJ, 519, 1

\bibitem[{Strateva {et~al.}(2001)}]{Strateva01}
Strateva, I. {et~al.} 2001, AJ, 122, 1861

\bibitem[{Tegmark {et~al.}(2004)Tegmark, Strauss, Blanton,
  {et~al.}}]{Tegmark04}
Tegmark, M., Strauss, M.~A., Blanton, M.~R., {et~al.} 2004, PhRvD, 69, 13501

\bibitem[{{Thompson}(2003)}]{Thompson03}
{Thompson}, R.~I. 2003, \apj, 596, 748

\bibitem[{{Thompson} {et~al.}(2006){Thompson}, {Eisenstein}, {Fan},
  {Dickinson}, {Illingworth}, \& {Kennicutt}}]{Thompson06}
{Thompson}, R.~I., {Eisenstein}, D., {Fan}, X., {Dickinson}, M., {Illingworth},
  G., \& {Kennicutt}, R.~C. 2006, astro-ph/0605060

\bibitem[{{Totani} {et~al.}(2001){Totani}, {Yoshii}, {Iwamuro}, {Maihara}, \&
  {Motohara}}]{Totani01a}
{Totani}, T., {Yoshii}, Y., {Iwamuro}, F., {Maihara}, T., \& {Motohara}, K.
  2001, \apjl, 550, L137

\bibitem[{Tresse \& Maddox(1998)}]{Tresse98}
Tresse, L. \& Maddox, S.~J. 1998, ApJ, 495, 691

\bibitem[{Tresse {et~al.}(2002)Tresse, Maddox, F\'{e}vre, \& Cuby}]{Tresse02}
Tresse, L., Maddox, S.~J., F\'{e}vre, O.~L., \& Cuby, J.-G. 2002, MNRAS, 337,
  369

\bibitem[{Tully {et~al.}(1998)Tully, Pierce, J.-S.~Huang, Verheijen, \&
  Witchalls}]{Tully98}
Tully, B., Pierce, M.~J., J.-S.~Huang, W.~S., Verheijen, M. A.~W., \&
  Witchalls, P.~L. 1998, AJ, 115, 2264

\bibitem[{{Vale} \& {Ostriker}(2004)}]{Vale04}
{Vale}, A. \& {Ostriker}, J.~P. 2004, \mnras, 353, 189

\bibitem[{Williams {et~al.}(1996)}]{Williams96}
Williams, R.~E. {et~al.} 1996, AJ, 112, 1335

\bibitem[{York {et~al.}(2000)}]{York00}
York, D.~G. {et~al.} 2000, AJ, 120, 1579

\end{thebibliography}
\end{document}